%% file: ArXiv1.tex
\documentclass[11pt,a4paper]{article}
\usepackage{jheppub}
\usepackage{mathrsfs,graphicx,rotating,amsmath,amsfonts,mathtools,booktabs,wasysym,caption}
\usepackage{hyperref}
\usepackage{slashed}
\usepackage[table,xcdraw,dvipsnames]{xcolor}
\usepackage{graphicx}
\usepackage{bbold}
\usepackage[utf8x]{inputenc}
\usepackage[english]{babel}
\usepackage{multirow,multicol}
\usepackage{epstopdf}
\usepackage{bbm}
\usepackage{changepage}
\usepackage{appendix}
\usepackage{libertine}
\usepackage{braket}
\usepackage{wasysym}
\usepackage{empheq}
\usepackage{cancel}
\usepackage{enumitem}
\usepackage{mathrsfs}
\usepackage{lmodern}
\usepackage{tabularx}
\usepackage{multicol}
\usepackage{color}
\usepackage{mathtools}
\usepackage{verbatim}
\usepackage{amssymb}
\usepackage{amsfonts}

\setcitestyle{square}

\DeclarePairedDelimiter\abs{\lvert}{\rvert}%
\title{Exploring The Holographic Swampland} 
\author[a]{Joseph P. Conlon,}\author[a]{Sirui Ning,}\author[a]{Filippo Revello}
\affiliation[a]{Rudolf Peierls Centre for Theoretical Physics\\ Beecroft Building, Clarendon Laboratory, Parks Road, University of Oxford, OX1 3PU, UK}
\emailAdd{joseph.conlon@physics.ox.ac.uk}\emailAdd{sirui.ning@physics.ox.ac.uk}\emailAdd{filippo.revello@physics.ox.ac.uk}
\abstract{We extend studies of holographic aspects of moduli stabilisation scenarios to both fibred versions of LVS and the type IIA DGKT flux vacua. We study the holographic properties of the low-energy moduli Lagrangian that describes both the AdS vacuum and also small perturbations about it. For type IIA vacua in the large-volume regime, the CFT data (operator dimensions and higher-point interactions) take a universal form independent of the many arbitrary flux choices, as was previously found for LVS stabilisation. For these IIA vacua the conformal dimensions of the dual operators are also, surprisingly, all integers, although we do not understand a deeper reason why this is so. In contrast to behaviour previously found for LVS and KKLT, the fibred models also admit cases of mixed double-trace operators (for two different axion fields) where the anomalous dimension is positive.}
\begin{document}

\maketitle
\section{Introduction}\label{sc:intro}

If string theory is the true underlying theory of this world, then the observed universe exists as a vacuum state of string theory. This makes it important for physicists to understand the space of allowed vacua of string theory; in particular, vacua with three large spatial dimensions with all remaining dimensions compactified.

The traditional approach to understanding string compactifications was based on top-down analyses and constructions of string vacua (for reviews see Refs. \cite{Douglas:2006es,Gra_a_2006,Conlon:2006gv,Denef:2008wq}). This aimed at obtaining such 4-dimensional vacua by starting with the full 10-dimensional theory and then proceeding through a series of controlled approximations to emerge eventually with controlled 4-dimensional solutions. Typically, the progression was from 10-dimensional string theory to 10-dimensional supergravity to 4-dimensional supergravity effective field theories -- and then, finally, to a vacuum derived within this framework of 4-dimensional effective field theory.

The claim to have a controlled solution was provided in three ways. The first means of control was 
through working with approximately supersymmetric compactifications (typically with Calabi-Yau internal manifolds)
and having supersymmetry broken only at scales far below the compactification scale. This allowed higher-dimensional questions to remain within the framework of supersymmetry and benefit from the additional control provided. The second means of control was to insist that the internal manifold was in the geometric regime, so that the $\alpha'$ expansion was parametrically controlled. The final means of control was to work at small string coupling, thereby allowing control of the string loop expansion.

In terms of obtaining string vacua, these restrictions are (mostly) not ones of principle. They are imposed not because of any belief that vacua \emph{cannot} exist at strong coupling with badly-broken supersymmetry, but rather through the (correct) sense that it is extremely hard to \emph{establish in a convincing fashion} that such vacua exist.

While there are clear limitations in restricting ourselves to vacua which are at weak coupling, there is one important physical reason to focus on such vacua -- the fact that the observed universe, both in particle physics and cosmology, is characterised by small numbers and weak couplings. This statement becomes more, rather than less, true as we move to the highest energy scales. Examples are the inflationary density perturbations (characterised by $\frac{\delta \rho}{\rho} \sim 10^{-5}$), the gauge couplings of the Standard Model $\alpha_{SU(3)}$, $\alpha_{SU(2)}$ and $\alpha_{U(1)_Y}$ (which all become weakly coupled at high but sub-Planckian energies) or the Yukawa couplings of the Standard Model.

Recent years have seen an alternative approach to understanding string vacua: the Swampland program \cite{Vafa:2005ui,Ooguri:2006in,Palti:2019pca}. This aims at determining general methods to exclude many low-energy Lagrangians as possible vacua of string theory, by finding principles that must \emph{always} be satisfied within quantum gravity. In this way it can (potentially) bypass the many steps involved in more traditional approaches using dimensional reduction and effective field theory. While such bold conjectures offer hopes of circumventing the need for laborious construction and control of individual vacua, it replaces the hard task of ensuring parametric control of vacua with the equally hard task of proving, or at least establishing significant evidence for, such general conjectures.

Nonetheless -- new perspectives on old problems are still beneficial. This is one motivation for our work here, which is focused on analysing
`traditional' vacua from a holographic perspective. This has two purposes. The first is to determine whether the holographic perspective reveals any interesting structure within established models of moduli stabilisation which may be opaque in more traditional treatments. The second is to determine whether manifestly `swampland' modifications within a supergravity effective field theory can be related to defined modifications within a dual CFT. In an ideal world, this may provide a hope of understanding swampland constraints in terms of allowed behaviour within CFTs \cite{Conlon:2018vov,Conlon:2020wmc}. While CFT approaches to the swampland have already attracted a significant degree of attention in the literature \cite{Nakayama:2015hga,Benjamin:2016fhe,Montero:2016tif,Giombi:2017mxl,Urbano:2018kax,Harlow:2018tng,Harlow:2018jwu,Montero:2018fns,Baume:2020dqd,Perlmutter:2020buo,Kundu:2021qpi,Aharony:2021mpc,Antipin:2021rsh}, not much work has been carried out in the context of specific realisations.  There was some discussion in older literature about properties of duals to explicit moduli stabilisation constructions \cite{Silverstein:2003jp,Aharony:2008wz,deAlwis:2014wia}, but the description is very complex and many details still remain to be worked out. In this work, we shall take a different approach and exploit parametric limits (such as large volume) which allow one to isolate treatable, effective CFTs with a large gap to the heavier modes.

This paper is structured as follows. In section \ref{sc:lvs} we review some aspects of CFT and the computation of anomalous dimensions. In section \ref{sec:fibre} we discuss the case of fibred versions of LVS, where there are multiple light moduli compared to `standard' LVS, whereas in section \ref{sec:IIA} we discuss type IIA flux compactifications and the corresponding operator dimensions. In section \ref{sec:Conclusions} we conclude.

\section{CFT Preliminaries }\label{sc:lvs}
The subset of Conformal Field Theories (CFTs) with a weakly coupled gravity dual go by the name of Holographic CFTs, and display some peculiar properties which we attempt to summarize here. While much of the discussion carries over from \cite{Conlon:2020wmc}, this concise introduction aims to make the presentation as self-contained as possible.

In weakly coupled scenarios, correlation functions in AdS can be calculated using standard perturbation theory, and the results mapped to the CFT side with the aid of the holographic dictionary. For the former to be possible, both curvature and string coupling corrections must be under control: $R_{AdS} \gg \ell_S$ and $g_s \ll 1$. In the most celebrated incarnation of the correspondence - type IIB String Theory on $AdS_5 \times S^5$ dual to $\mathcal{N}=4$ SYM - this amounts to the requirement $N \gg 1$, since the t'Hooft coupling $\lambda = g^2 N$ is proportional to $(R/\ell_s)^d$. More generally, Holographic CFTs can be characterized by the following two properties:
\begin{itemize}
\item A finite number of primary operators with conformal dimensions of $\mathcal{O}(1)$, which are separated from the rest of the spectrum by a large gap $\Delta_{\text{gap}} \gg 1$.
\item The existence of an expansion parameter playing the role of large-$N$, around which a perturbative series can be organised. By analogy, this is always denoted by $N$ in the literature.
\end{itemize}
It has also been conjectured that these property are both necessary and sufficient conditions for a CFT to be holographic \cite{Heemskerk:2009pn}.
At the first non-trivial order in $1/N$, the operator content of such theories consists of the primaries $\{ \mathcal{O}_i \} $ and the double trace operators,\footnote{In a realistic setup, one of course also has to include the stress energy tensor $T_{\mu \nu}$ alongside any double trace operators that can be constructed using it.} which are (schematically) built out of the former as 
\begin{equation}
[\mathcal{O}_i \mathcal{O}_j]_{n,\ell} \equiv \mathcal{O}_i \Box^n \partial_{\mu_1} ... \partial_{\mu_\ell} \mathcal{O}_j.
\end{equation}
For each value of $n$ and $\ell$, only one such operator exists. Their conformal dimension is given by
\begin{equation}
\Delta_{n,\ell} = \Delta_i + \Delta_j + 2n + \ell + \gamma(n,\ell),
\end{equation}
which is the sum of the classical dimension, $\Delta_i + \Delta_j + 2n + \ell$, and the anomalous dimension $\gamma(n,\ell)$ of magnitude $\mathcal{O}(1/N^2)$. Alongside the OPE coefficients, the anomalous dimensions (which can be viewed as the binding energy in AdS space) encode all the dynamical information about the theory and will be central quantities in our analysis. In principle they can be extracted from the expression of AdS 4-point correlators or using a Bootstrap approach, but the procedure is not always straightforward. In section \ref{ssc:ls}, we will however show how they can be systematically computed in the particular case that we will be concerned with, i.e. anomalous dimensions $\gamma(n,\ell)$ with $n =0$. The relevance of such a scenario is explained in the next section, in the context of CFT positivity bounds.

\subsection{Positivity Bounds}\label{ssc:pb}
The requirements of conformal and crossing symmetry alone, cornerstones 
of the Bootstrap program \cite{Ferrara:1973yt,Polyakov:1974gs,Mack:1975jr}, are not sufficient to impose any non-trivial bounds on the OPE data for holographic CFTs. Indeed, it was shown in \cite{Heemskerk:2009pn} how all the higher order operator coefficients in the Lagrangian of a single scalar on AdS map to a single solution of the bootstrap equations. On the other hand, the physical principles of causality and unitarity often impose additional constraints in the space of possible theories, and CFTs make no exception. By demanding that every state in a conformal multiplet has a non-negative norm, for example, one is led to the following bounds for the conformal dimensions of primary operators \cite{Mack:1975je,Minwalla:1997ka}:
\begin{equation}
\Delta \geq \frac{d-2}{2} \quad  {\rm{for}}  \,\,  \ell = 0, \quad \quad \quad \Delta \geq \ell +d-2 \quad  {\rm{for}} \,\,  \ell > 0
\end{equation}
More elaborate consequences of these fundamental principles have been pioneered in \cite{Komargodski:2012ek,Fitzpatrick:2012yx}, resulting in the formulation of positivity bounds for towers of operators with minimal twist.\footnote{The twist of an operator is denoted by is defined as $\tau = \Delta - \ell$.}  Heuristically, the latter give the dominant contribution to the Lorentzian OPE in certain kinematic regimes, and thus can be constrained more easily. More precisely, if one denotes by $\tau^*(\ell)$ the twist of the minimal twist operators with spin $\ell$ appearing in the OPE $\mathcal{O} \mathcal{O}^{\dag}$ of a scalar with itself, the following bound applies for $\ell$ even and larger than a critical value $\ell_2 \geq 2$ (with $l_3 > l_2$):
\begin{equation}\label{eq:pos}
\frac{\tau^*(\ell_3)-\tau^*(\ell_1)}{\ell_3-\ell_1} \leq \frac{\tau^*(\ell_2)-\tau^*(\ell_1)}{\ell_2-\ell_1}.
\end{equation}
Although the authors of Refs \cite{Komargodski:2012ek,Fitzpatrick:2012yx} had to invoke additional assumptions on the behaviour of off-shell amplitudes, these were later dropped in subsequent proofs \cite{Costa:2017twz}, where the result was also generalized to any $\ell >1$ through the OPE inversion formula of \cite{Caron-Huot:2017vep}.  In cases where gravity can be decoupled -- i.e. an effective theory on AdS characterized by a cutoff scale $\Lambda_c \ll M_{P}$ -- the leading twist operators in the $\mathcal{O} \mathcal{O}^{\dag}$ OPE are the double traces $[\mathcal{O}\mathcal{O} ]_{0,\ell}$, and (\ref{eq:pos}) can be turned into a convexity statement on their anomalous dimension:
\begin{equation}\label{eq:pos}
\frac{\gamma(0,\ell_3)-\gamma(0,\ell_1)}{\ell_3-\ell_1} \leq \frac{\gamma(0,\ell_2)-\gamma(0,\ell_1)}{\ell_2-\ell_1}.
\end{equation}
Given that the anomalous dimensions asymptote to zero as $\ell \rightarrow \infty$, they will also have to be negative for $\ell >1$. In a different context, similar results concerning the negativity of the $\gamma(0,2)$ anomalous dimension were also proven in \cite{Hartman:2015lfa}. More recently, such convexity relations have also been generalized to treat the OPE of two non-identical scalar primaries \cite{Kundu:2020gkz} and other closely related bounds for EFTs on AdS (which include the effect of dynamical gravity) have been obtained in \cite{Kundu:2021qpi,Caron-Huot:2021enk}.

These results served as a partial motivation in our previous paper \cite{Conlon:2020wmc} to conjecture that a similar negativity property for anomalous dimensions would correlate with Swampland conditions on AdS, supported from a number of examples in 4d string compactifications.\footnote{The original condition (\ref{eq:pos}) leads to trivial constraints for the type of Lagrangian we considered \cite{Conlon:2020wmc} .} In particular, we proposed that the mixed anomalous dimensions $\gamma^{\rm{mix}}_{12}(0,\ell)$, corresponding to the double trace operators of non-identical scalars $[\mathcal{O}_1\mathcal{O}_2 ]_{0,\ell}$ also be negative in CFTs dual to scenarios of moduli stabilization (based on results for racetrack, LVS and KKLT stabilisation scenarios). For simplicity, we considered anomalous dimensions for $\ell \gg1$, which are not affected by contact terms in the lagrangian and can be easily computed using a formula shown in the next section. The rest of this paper is devoted to the exploration of this condition in a different class of scenarios. 

\subsection{Anomalous Dimensions at Large Spin}\label{ssc:ls}
The purpose of this section is to show explicitly how large spin anomalous dimensions of the form $\gamma(0,\ell)$ can be calculated from the effective action on AdS. Although there are no general, analytic formulas to extract anomalous dimensions for any $(n,\ell)$ and arbitrary values of the external dimensions, the case we are interested is rather special. In particular, Ref. \cite{Costa:2014kfa} derived the formal expression
\begin{equation}\label{eq:adf}
\begin{split}
\gamma^{\rm{mix}}_{12}(0,\ell)  = - \int_{-i \infty}^{+ i \infty} & \frac{dt}{2 \pi i } M_{12 \rightarrow 12}(0,t)\,\, _3F_2(-\ell, \Delta_1+\Delta_2 + \ell-1,\frac{t}{2}; \Delta_1, \Delta_2;1 ) \\ & \times \Gamma \Big(\Delta_1-\frac{t}{2}\Big)\, \Gamma \Big(\Delta_2-\frac{t}{2} \Big)\, \Gamma \Big(\frac{t}{2}\Big)^2,
\end{split}
\end{equation}
where $M(s,t)_{12 \rightarrow 12}$ is the Mellin amplitude corresponding to the $\langle \mathcal{O}_{1} \mathcal{O}_{2} \mathcal{O}_1 \mathcal{O}_2 \rangle$ correlator.\footnote{ We follow the kinematic conventions of \cite{Costa:2014kfa}, which differ from those adopted in \cite{Conlon:2020wmc} by an exchange of the s- and t-channels.} Mellin amplitudes \cite{Mack:2009mi,Mack:2009gy,Penedones:2010ue,Penedones:2016voo} are particular representations of conformal correlators which most closely resemble scattering amplitudes in flat spacetime, and in this case $M(s,t)_{12 \rightarrow 12}$ is the analogue of a $1 2 \rightarrow 1 2$ scattering amplitude. For a fuller development of Mellin amplitudes in the context of AdS moduli stabilisation scenarios, we refer readers to our earlier work \cite{Conlon:2020wmc}.

Using (\ref{eq:adf}), we will later see that only exchange diagrams contribute to the sum at leading order in the $1/\ell$ expansion, and in particular only in the t-channel. Therefore, the final result only depends on certain cubic couplings. Contact terms, for example, contribute up to a finite spin $\ell_f  = N_{\partial}/2$, where $N_{\partial}$ is equal to number of derivatives it contains. In particular, the exchange of a given primary operator $\mathcal{O}$ induces an anomalous dimension for the mixed double trace operators $[\mathcal{O}_1 \mathcal{O}_2]_{0,\ell}$ of
\begin{equation}\label{eq:adim}
\gamma^{\rm{mix}}_{12}(0,\ell) \underset{\ell \gg 1}{\simeq}  -\frac{2 \Gamma(\Delta_{\mathcal{O}})}{\Gamma(\frac{\Delta_{\mathcal{O}}}{2})^2} \frac{\Gamma(\Delta_1) \Gamma(\Delta_2)}{\Gamma(\Delta_1-\frac{\Delta_{\mathcal{O}}}{2})\Gamma(\Delta_2-\frac{\Delta_{\mathcal{O}}}{2})} \frac{C_{11\mathcal{O}} C_{22\mathcal{O}}}{\ell^{\Delta_{\mathcal{O}}}},
\end{equation}
where $C_{11\mathcal{O}},C_{22\mathcal{O}}$ are the coefficients of the three-point functions $\langle \mathcal{O}_{1(2)} \mathcal{O}_{1(2)} \mathcal{O} \rangle$.
Although the same result could have been more easily obtained by solving the bootstrap crossing equation at leading order \cite{Li:2015rfa}, we emphasize that formula (\ref{eq:adf}) also provides a systematic way to expand the answer in subsequent powers of $1/ \ell$, and eventually resum the series to obtain the finite spin result. It also allows one to keep the analogy with positivity bounds from flat space more manifest, given the similarity of Mellin amplitudes to scattering amplitudes in flat spacetime. 

The three-point function coefficients in (\ref{eq:adim}) can be computed by evaluating the corresponding Witten diagram in AdS, which is most easily performed in the embedding formalism \cite{Penedones:2010ue,Penedones:2016voo}. Given the particular structure of the Lagrangians we are going to consider, the cubic coupling will always be of the general form
\begin{equation}
\mathcal{L} \supset c_1 \varphi_A^2 \varphi_B + c_2 \partial_{\mu} \varphi_C \partial_{\mu} \varphi_C \varphi_D.
\end{equation}
Then, the corresponding coefficients take the form \cite{Conlon:2005ki}
\begin{equation}\label{eq:opec1}
C_{A A B} = c_1(2+4\delta_{AB})\frac{ \pi^{d/2}  \Gamma \big( \frac{2 \Delta_{A}+\Delta_B-d}{2} \big)}{ \Gamma (\Delta_{A})^2\Gamma (\Delta_{B})}.
\end{equation}
\begin{equation}\label{eq:opec2}
C_{C C D } = c_2 \frac{ \pi^{d/2}  \Gamma \big( \frac{2 \Delta_C+ \Delta_{D}-d}{2} \big)}{\Gamma (\Delta_{C})^2 \Gamma (\Delta_D)}  (\Delta_C^2-2\Delta_{C}-  \Delta_D+3).
\end{equation}
\subsection{A Special Case: Degenerate Conformal Dimensions}\label{ssc:deg}
When a CFT contains two primaries with integer spaced conformal dimensions, the spectra of the corresponding double trace operators overlap, and one needs to be more careful in the computation of the anomalous dimensions. To be explicit, let us consider two scalars $\mathcal{O}_1$ and $\mathcal{O}_2$ with scaling dimensions $\Delta_1$, $\Delta_2$. At zeroth order in the large-$N$ expansion, the double trace operators $[\mathcal{O}_1 \mathcal{O}_1]_{n_1,\ell}$ and $[\mathcal{O}_2 \mathcal{O}_2]_{n_2,\ell}$ will be degenerate if the dimensions satisfy 
\begin{equation}\label{eq:degd}
2(n_1 -n_2)= \Delta_2 -\Delta_1, \quad \quad n_1,n_2 \in \mathbb{N}
\end{equation}
and analogous conclusions hold with the mixed operators $[\mathcal{O}_1 \mathcal{O}_2]_{n,\ell}$ too. While from an AdS perspective Eq. (\ref{eq:degd}) would seemingly correspond to a fine-tuning conspiracy on the value of the masses,\footnote{In the type IIA models of section \ref{sec:IIA} we will surprisingly find examples where this condition is satisfied. However, we expect it to be broken by subleading volume effects.} there is one notable exception where this condition naturally occurs, namely when two fields are massless and $\Delta_1 = \Delta_2 = d$. One such example is provided by the case where the potential has flat directions corresponding to certain moduli or, alternatively, for axions protected by shift invariance. In large volume models, the latter scenario is quite robust, since any additional non-perturbative effects lifting the axion mass are suppressed as $m \propto e^{- b\mathcal{V}^{2/3}}$, which does not appear at any order in a $1/N$ expansion.\footnote{Since the volume plays the role of large-$N$ in such models.}

Of course, in any realistic stringy example -- such as those we will consider in sections \ref{sec:fibre} and \ref{sec:IIA} -- particle masses (and hence the dimensions of primaries) may receive higher order contributions that break the degeneracy. Since we are working in a perturbative framework, this is only relevant if the corrections to the conformal dimensions dominate parametrically over the anomalous dimensions, i.e. they have to be strictly larger than $\mathcal{O}(1/N^2)$. If this happens for $\Delta_1 \neq \Delta_2 $, the operator mixing is effectively eliminated, and anomalous dimensions can be calculated as usual. On the other hand, if $\Delta_1 = \Delta_2 $ the degeneracy is still resolved in principle (unless the symmetries of the problem conspire to leave the degeneracy still unbroken at this subleading order), but mixing effects can appear between $\mathcal{O}_1$ and $\mathcal{O}_2$. Indeed, the true primaries are the eigenstates of the degeneracy-breaking perturbation, which can be rotated from the original fields by an unknown angle.\footnote{Crucially, this angle remains finite if the magnitude of the perturbation approaches zero.} Therefore, the form of the perturbation becomes necessary to calculate the anomalous dimensions even at leading order, and in such cases we will not be able to provide an expression for $\gamma(0,\ell)$. Otherwise, the operators should effectively be considered degenerate, and it will be necessary to use the techniques outlined in this section.

\subsubsection{Mixing Matrix}\label{ssc:ofpb}
For concreteness, let us focus on the case where $\Delta_1 = \Delta_2 \equiv \Delta $ and no degeneracy breaking is present. For each given pairing $(n,\ell)$, there is a three dimensional degenerate space, corresponding to the operators 
\begin{equation}\label{eq:multop}
[\mathcal{O}_1 \mathcal{O}_1]_{n,\ell}\, ; \,\, [\mathcal{O}_2 \mathcal{O}_2]_{n,\ell} \, ; \,\,  [\mathcal{O}_1 \mathcal{O}_2]_{n,\ell} .
\end{equation} 
They will be denoted by the double indices $[i j]$, taking values in ${[11],[22],[12]}$. In this subspace, the dilatation operator is diagonal only at order zero in the $1/N$ expansion, and the scaling dimensions are given by its eigenvalues: this is totally analogous to what happens in standard QM perturbation theory. In the approach of \cite{Fitzpatrick:2010zm}, anomalous dimensions are calculated by applying QM perturbation theory to the dilatation operator, and in a non-degenerate system this amounts to evaluating the expectation value of the interaction term $\Delta \hat{D}$ on a given state. In a degenerate system, one instead has to diagonalize
\begin{equation}
\bar{\gamma}_{[ij]\,[kl]}(n,\ell)= \,_{ij}\bra{n,\ell} \Delta \hat{D} \ket{n,\ell}_{k l}  
\end{equation}
where $\ket{n,\ell}_{i j} $ is the state created by acting with $[\mathcal{O}_i\mathcal{O}_j]_{n,\ell}(0)$ on the vacuum, and the bar is used to distinguish between the matrix elements and the actual anomalous dimensions, given by the eigenvalues. The eigenvectors of the matrix determine the basis of double trace operators in which the dilatation operator is diagonal at the first non-trivial order in $1/N$,  as a linear combination of the three operators in (\ref{eq:multop}). 

Although our approach to the evaluation of large spin anomalous dimensions is technically (and conceptually) different, anomalous dimensions can still be obtained as the eigenvalues of an analogous mixing matrix $ \tilde{\gamma}_{[ij]\,[kl]}(0,\ell)$. In particular, we show in Appendix \ref{appendix:deg} that the matrix elements represent the ``effective" anomalous dimension extracted from the mixed correlators
\begin{equation}
\langle \mathcal{O}_i\mathcal{O}_j\mathcal{O}_k\mathcal{O}_l \rangle \supset  \frac{f_0(n,\ell)^2\tilde{\gamma}_{[ij][kl]}(n,\ell)}{N^2} \partial_{\Delta} G_{2\Delta+2n+\ell,\ell},
\end{equation}
where the $f_0(0,\ell)$ are the generalized free theory OPE coefficients. They can be calculated with the analogue of Eq.(\ref{eq:adim}) for the Mellin amplitude corresponding to the $\langle \mathcal{O}_{i} \mathcal{O}_{k} \mathcal{O}_j \mathcal{O}_l \rangle$  correlator.

If we assume that at least one of the two primaries is odd with respect to a $\mathbb{Z}_2$ symmetry, as is the case for operators dual to an axion on AdS, the system becomes two-dimensional. This is the only case we will encounter in practice, as the degeneracy will always be broken in other systems. In this case, only the $[\mathcal{O}_1 \mathcal{O}_1]^i_{n,\ell} $ and $[\mathcal{O}_2 \mathcal{O}_2]^i_{n,\ell} $ operators have the same transformation property under the discrete symmetry and can mix amongst themselves. In this particular case, the eigenvalues take the simplified form
\begin{equation}\label{eq:ev}
\gamma_{\pm} = \frac{\tilde{\gamma}_{[11][11]}+\tilde{\gamma}_{[22][22]}\pm \sqrt{ (\tilde{\gamma}_{[11][11]}-\tilde{\gamma}_{[22][22]})^2+4 \tilde{\gamma}_{[11][22]}^2 }} {2},
\end{equation}
where the indices $n,\ell$ have been suppressed for simplicity. An interesting consequence of the modified formula (\ref{eq:ev}) is that one of the anomalous dimensions can apparently be positive,\footnote{Since $\tilde{\gamma}_{[11][11]}$ and $\tilde{\gamma}_{[22][22]}$ will still be positive by construction.} when $\tilde{\gamma}^2_{[11][22] } > \tilde{\gamma}_{[11][11]} \tilde{\gamma}_{[22][22]}$.
Even though we are dealing with identical operators, there does not seem any reason why this behaviour should be inconsistent.
 The reason is that in the OPE of two identical primaries, only the lowest twist operators have to obey a convexity condition (implying negativity) - therefore it is enough to have just one double trace operator with negative anomalous dimension appearing for every value of $\ell$. For instance, the OPE of $\mathcal{O}_1$ with itself will contain two classes of double trace operators, with positive and negative anomalous dimensions respectively. The lowest twist operators will be those belonging to the latter class, and these satisfy the required condition irrespective of the existence of those with a positive anomalous dimension.   

\section{Fibred Calabi-Yau Scenarios of Moduli Stabilization}\label{sec:fibre}

In our earlier work \cite{Conlon:2020wmc} we studied holographic aspects of certain moduli stabilisation scenarios, in particular LVS, KKLT and racetrack scenarios. These displayed interesting properties. LVS displayed a universality of conformal dimensions in the large-volume limit, and all scenarios had an interesting negative behaviour for the mixed anomalous dimensions. We are therefore interested more scenarios, and in this section we extend these studies to fibred versions of the original LVS scenario.

\subsection{A Brief Review of Fibred Calabi-Yau Scenarios}

One interesting variation on the original LVS is to consider models with fibred Calabi-Yau (CY) structures. In such models, there are multiple large moduli associated both to base and fibre moduli \cite{Cicoli:2008gp,Cicoli:2011it}. We shall take the form of the volume of a fibred Calabi-Yau to be given as the following (based on \cite{Cicoli:2008gp}):
\begin{equation}
\mathcal{V}=f_{\frac{3}{2}}(\tau_{j})-\sum_{i=1}^{N_{small}}\lambda_{i}\tau_{i}^{3/2},
\end{equation} 
where $f$ is a homogeneous function of degree 3/2, $j$ runs from 1 to $N_{large}$ and $h_{1,1}=N_{small}+N_{large}$, where $N_{large}$ is the number of large moduli and $N_{small}$ is the number of blow up moduli.

The fibred CY shares a similar feature with the ``Swiss Cheese'' form used in the original LVS, in terms of the blow-up moduli included in the volume. The difference lies in the function $f$, as this now involves multiple moduli: both the overall volume modulus and the fibration moduli. Different moduli play different roles. The volume modulus controls the LVS potential. The fibre modulus is a flat direction of the LVS potential, which means it will be absent in the LVS potential at leading order. However, it will appear in the kinetic terms, and also at subleading order in the string loop expansion. The blow up modulus controls the heavy mode which will subsequently be integrated out of the LVS potential within the low-energy effective field theory.

In the next two subsections, we will use two specific examples to illustrate these ideas which will involve respectively one or two fibre moduli.

\subsection{ Fibred Calabi-Yau: Example with One Fibre Modulus}

We will first look at moduli stabilisation in order to establish the AdS effective action (essentially a review of \cite{Gra_a_2006}), before considering the holographic implications of this effective action.

\subsubsection{Moduli Stabilisation}

We assume as in \cite{Cicoli:2008gp} that the volume takes the form: 
\begin{equation}
\mathcal{V}=\alpha(\sqrt{\tau_{1}}\tau_{2})-\gamma\tau_{3}^{3/2}.
\end{equation}
 Here $\tau_{1}$ controls the volume of the fibre, $\tau_{2}$ controls the volume of the base and $\tau_{3}$ describes the blow up modulus.

The Large Volume Scenario is a specific type IIB flux compactification moduli stabilisation scenario, which stabilises all moduli in a non-SUSY AdS vacuum at an exponentially large volume $\mathcal{V}$ (as per \cite{Balasubramanian:2005zx, Conlon:2005ki,Conlon:2006gv}). Following the usual approach in LVS, after stabilising  and integrating out the complex structure moduli and dilaton,  the K\" ahler potential and superpotential read:
\begin{equation}
K=-2 \textrm{ln}(\mathcal{V}+\hat{\xi}), 
\end{equation}
\begin{equation}
W=W_{0}+A_{3}e^{-a_{3}T_{3}},
\end{equation}
\begin{equation}
\hat{\xi}=\frac{\xi}{g_{s}^{\frac{3}{2}}},
\end{equation}
\begin{equation}
\xi=-\frac{\chi(X)\zeta(3)}{2(2\pi)^{3}},
\end{equation}
Here $\chi(X)$ is the Euler number of the Calabi-Yau manifold X,  $\zeta$ is the Riemann zeta function with $\zeta(3)\approx1.2$. $T_{3}$ is the chiral multiplet for the blow-up K\" ahler modulus, $T_{3}=\tau_{3}+iC_{3}$.

The $\mathcal{N}=1$ F-term supergravity scalar potential takes the following form:
\begin{equation}
V=e^{K}(K^{i\overline{j}}D_{i}WD_{\overline{j}}\overline{W}-3|W|^{2}),
\end{equation}
where we have set $M_{P}=1$. After substituting the K\" ahler potential and superpotential into the scalar potential, the effective potential becomes:

\begin{equation}
V_{scalar}=\frac{Aa_{3}^{2}\sqrt{\tau_{3}}}{\mathcal{V}}e^{-2a_{3}\tau_{3}}-\frac{BW_{0}a_{3}A_{3}\tau_{3}}{\mathcal{V}^{2}}e^{-a_{3}\tau_{3}}+\frac{C\xi W_{0}^{2}}{g_{s}^{\frac{3}{2}}\mathcal{V}^{3}},
\end{equation}
where $A,B,C$ are numerical constants. The important point is that the fibre modulus does not appear in the effective potential at tree level, and so at this order is a flat direction of the scalar potential. Therefore there is an additional light scalar degree of freedom, which is an interesting feature of fibred Calabi-Yau models compared to the original LVS. 

As with standard LVS, this scalar potential has a minimum at exponentially large volumes,
\begin{equation}
\langle \tau_{3}\rangle\sim \frac{\xi^{\frac{2}{3}}}{g_{s}},
\end{equation}
\begin{equation}
\langle \mathcal{V} \rangle\sim e^{a_{3}\langle\tau_{3}\rangle}.
\end{equation}

After integrating out the blow up modulus $\tau_{3}$, the effective potential for the LVS volume modulus reads \cite{Conlon:2018vov}:
\begin{equation}
V_{scalar}=[-A^{'}\left(\textrm{ln}(f\mathcal{V})\right)^{\frac{3}{2}}+B^{'}]\frac{W_{0}^{2}}{\mathcal{V}^{3}}.
\end{equation}
where $A^{'}$, $B^{'}$ are numerical constants.

\subsubsection{Kinetic Terms and Canonical Fields}
The effective field theory of these fibred models contains four light moduli: one overall volume modulus which is massive, plus one fibre modulus and two axions, which are approximately massless.
In order to see how they interact with each other, we write down the kinetic terms for the theory in the large volume limit using standard EFT techniques. Neglecting higher order terms, the kinetic terms read:
\begin{equation}
L_{kin}=\frac{1}{4\tau_{1}^{2}}\partial_{u}\tau_{1}\partial^{u}\tau_{1}+\frac{1}{2\tau_{2}^{2}}\partial_{u}\tau_{2}\partial^{u}\tau_{2}
+\frac{1}{4\tau_{1}^{2}}\partial_{u}a_{1}\partial^{u}a_{1}+\frac{1}{2\tau_{2}^{2}}\partial_{u}a_{2}\partial^{u}a_{2}.
\end{equation}
As a first step in diagonalising the action, we re-express the mode $\tau_{2}$  in terms of the overall LVS volume:
\begin{equation}
\tau_{2}=\frac{\mathcal{V}}{\alpha\sqrt{\tau_{1}}}.
\end{equation}
The kinetic terms then become
\begin{equation}
\mathcal{L} \supset \frac{3}{8\tau_{1}^{2}}\partial_{u}\tau_{1}\partial^{u}\tau_{1}-\frac{1}{2\tau_{1}\mathcal{V}}\partial_{u}\tau_{1}\partial^{u}\mathcal{V}+\frac{1}{2\mathcal{V}^{2}}\partial_{u}\mathcal{V}\partial^{u}\mathcal{V}
+\frac{1}{4\tau_{1}^{2}}\partial_{u}a_{1}\partial^{u}a_{1}+\frac{\alpha^{2}\tau_{1}}{2\mathcal{V}^{2}}\partial_{u}a_{2}\partial^{u}a_{2}.
\end{equation}

We next diagonalise both kinetic terms and mass terms in order to get a canonically normalised theory.
\begin{equation}
\textrm{ln}\, \tau_{1}=a\Phi_{1}+b\Phi_{2},
\end{equation}
\begin{equation}
\textrm{ln}\, \mathcal{V} =c\Phi_{1}+d\Phi_{2}.
\end{equation}
Here $\Phi_{1}$ and $\Phi_{2}$ will be the canonically normalised fields. Before we go into the details of the calculation, we make an observation. As the mass term only depends on the overall LVS volume, we must choose either $c=0$ or $d=0$ when diagonalising. Here we choose $d=0$, so our ansatz becomes:
\begin{equation}
\begin{aligned}
&\textrm{ln}\,\tau_{1}=a\Phi_{1}+b\Phi_{2},\\
&\textrm{ln}\, \mathcal{V}=c\Phi_{1}.\\
\end{aligned}
\end{equation}

We refer interested readers to appendix \ref{app:1m} for the details. Here we just state the final results:

\begin{equation}
\begin{aligned}
&\textrm{ln}\, \tau_{1}=\sqrt{\frac{2}{3}}(\Phi_{1}+\sqrt{2}\Phi_{2}),\\
&\textrm{ln}\, \mathcal{V}=\sqrt{\frac{3}{2}}\Phi_{1}.\\
\end{aligned}
\end{equation}
The kinetic terms then read:
\begin{equation}\label{eq1}
\mathcal{L} \supset\frac{1}{2}(\partial_{u}\Phi_{1})^{2}+\frac{1}{2}(\partial_{u}\Phi_{2})^{2}+\frac{1}{2}e^{-2\sqrt{\frac{2}{3}}(\Phi_{1}+2\Phi_{2})}\partial_{u}a_{1}\partial^{u}a_{1}
+\frac{1}{2}e^{\left(\sqrt{\frac{2}{3}}-2\sqrt{\frac{3}{2}}\right)\Phi_{1}+2\sqrt{\frac{2}{3}}\Phi_{2}}\partial_{u}a_{2}\partial^{u}a_{2},
\end{equation}
where the axions have also been rescaled by a constant factor to achieve canonical normalization. The potential only depends 
on $\Phi_{1}$ and so is diagonal in this basis. Having expressed the Lagrangian in canonical form,  we can now move to a discussion of the interactions and their holographic interpretation.
\subsubsection{Holographic Interpretation}
From the couplings in (\ref{eq1}) and in the scalar potential, we can easily read off the OPE coefficients following Eqs (\ref{eq:opec1})-(\ref{eq:opec2}). The results are the following:
\begin{equation}
\begin{aligned}
&f_{a_{1}a_{1}}^{\Phi_{1}}= \sqrt\frac{2}{3} \frac{\pi^{\frac{d}{2}}\Gamma\left(\frac{2\Delta_{a_{1}}+\Delta_{\Phi_{1}}-d}{2}\right)}{\Gamma\left(\Delta_{\Phi_{1}}\right)\Gamma\left(\Delta_{a_{1}}\right)^{2}}\left(\Delta_{\Phi_{1}}+2\Delta_{a_{1}}-\Delta_{a_{1}}^{2}-3\right),\\
&f_{a_{1}a_{1}}^{\Phi_{2}}= 2\sqrt\frac{2}{3} \frac{\pi^{\frac{d}{2}}\Gamma\left(\frac{2\Delta_{a_{1}}+\Delta_{\Phi_{2}}-d}{2}\right)}{\Gamma\left(\Delta_{\Phi_{2}}\right)\Gamma\left(\Delta_{a_{1}}\right)^{2}}\left(\Delta_{\Phi_{2}}+2\Delta_{a_{1}}-\Delta_{a_{1}}^{2}-3\right),\\
&f_{a_{2}a_{2}}^{\Phi_{1}}=  \left( \sqrt{\frac{3}{2}} -\frac{1}{2}\sqrt{\frac{2}{3}} \right) \frac{\pi^{\frac{d}{2}}\Gamma\left(\frac{2\Delta_{a_{2}}+\Delta_{\Phi_{1}}-d}{2}\right)}{\Gamma\left(\Delta_{\Phi_{1}}\right)\Gamma\left(\Delta_{a_{2}}\right)^{2}}\left(\Delta_{\Phi_{1}}+2\Delta_{a_{2}}-\Delta_{a_{2}}^{2}-3\right),\\
&f_{a_{2}a_{2}}^{\Phi_{2}}= -\sqrt\frac{2}{3} \frac{\pi^{\frac{d}{2}}\Gamma\left(\frac{2\Delta_{a_{2}}+\Delta_{\Phi_{2}}-d}{2}\right)}{\Gamma\left(\Delta_{\Phi_{2}}\right)\Gamma\left(\Delta_{a_{2}}\right)^{2}}\left(\Delta_{\Phi_{2}}+2\Delta_{a_{2}}-\Delta_{a_{2}}^{2}-3\right),\\
&f_{\Phi_{1}\Phi_{1}}^{\Phi_{1}}=\frac{81\sqrt{6}}{2}\frac{3\pi^{\frac{d}{2}}\Gamma\left(\frac{3\Delta_{\Phi_{1}}-3}{2}\right)}{\Gamma\left(\Delta_{\Phi_{1}}\right)^{3}}.
\end{aligned}
\end{equation}

The key difference between the fibred Calabi-Yau LVS model and the original LVS model is that there are a total of 3 massless moduli in the effective field theory rather than just one. This leads to operator mixing in the operator product expansion. On the other hand, the volume modulus behaves exactly as in the pure LVS scenario, obtaining a conformal dimension of \cite{Conlon:2018vov,Conlon:2020wmc}
\begin{equation}
\Delta_{\Phi_1}= \frac{3(1+\sqrt{19})}{2},
\end{equation}
and with the same self-couplings.

In principle, the following double trace operators are degenerate
\begin{equation}
[a_{1}a_{1}]_{n,\ell},\quad [a_{2}a_{2}]_{n,\ell},\quad[a_{1}a_{2}]_{n,\ell},\quad [\Phi_{2}\Phi_{2}]_{n,\ell},\quad[a_{1}\Phi_{2}]_{n,\ell},\quad[a_{2}\Phi_{2}]_{n,\ell},
\end{equation}
while 
\begin{equation}
\quad[a_{1}\Phi_{1}]_{n,\ell}, \quad[a_{2}\Phi_{1}]_{n,\ell}
\end{equation}
also mix within themselves.
Each axion is invariant under an independent discrete $\mathbb{Z}_{2}$ symmetry acting as $a_i \rightarrow - a_i$, which results in selection rules for the double trace operators that can appear in a given OPE. This implies that states in different symmetry classes cannot mix with one another even though their classical dimension is the same. This is shown in Table \ref{tab:Z2}.

\begin{table}[h!]
\centering
\begin{tabular}{|c|c|c|c|}
\hline  
 &$\mathbb{Z}_{2}(a_{1})$&$\mathbb{Z}_{2}(a_{2})$\\
\hline 
$[a_{1}\Phi_{2}]_{n,\ell}$&-1&1\\
\hline 
$[a_{2}\Phi_{2}]_{n,\ell}$&1&-1\\
\hline 
$[a_{1}a_{2}]_{n,\ell}$&-1&-1\\
\hline 
$[a_{1}a_{1}]_{n,\ell}, [a_{2}a_{2}]_{n,\ell},[\Phi_{2}\Phi_{2}]_{n,\ell}$&1&1\\
\hline \hline
$[a_{1}\Phi_{1}]_{n,\ell}$&-1&1\\
\hline
$[a_{2}\Phi_{1}]_{n,\ell}$&1&-1\\
\hline
\end{tabular}
\caption{Irreducible representation of the double trace operators under the $\mathbb{Z}_{2} \times \mathbb{Z}_{2} $ symmetry. The two different blocks correspond to distinct subspaces, where the degeneracies are partly broken by the discrete symmetries.}
\label{tab:Z2}
\end{table}

For the non-degenerate operators involving different fields $ [a_{1}\Phi_{1}]_{0,\ell}, [a_{2}\Phi_{1}]_{0,\ell} 
 [a_{1}\Phi_{2}]_{0,\ell},$ $ [a_{2}\Phi_{2}]_{0,\ell}, [a_{1}a_{2}]_{0,\ell} $, we can directly refer to (\ref{eq:adim}) to work out the anomalous dimension. In particular, since the kinematic factors are equal for the two axions
\begin{equation}
\begin{aligned}
& \gamma(0,\ell)_{a_{1}\Phi_{1}}  \propto -f_{a_{1}a_{1}\Phi_{1}}f_{\Phi_{1} \Phi_{1} \Phi_{1}} < 0,\\
&\gamma(0,\ell)_{a_{2}\Phi_{1}}  =0,\\
&\gamma(0,\ell)_{a_{1}\Phi_{2}}  =0,\\
&\gamma(0,\ell)_{a_{2}\Phi_{2}} =0.\\
&\gamma(0,\ell)_{a_{1}a_{2}}\propto -f_{a_{1}a_{1}\Phi_{2}}f_{a_{2}a_{2}\Phi_{2}} >0,\\
\end{aligned}
\end{equation}
The first two anomalous dimensions are exactly those of LVS without fibred directions, and their agreement with our original conjecture should not come as a surprise. However, $\gamma(0,\ell)_{a_{1}a_{2}}$ is positive, violating the expected behaviour. The interpretation for this is that maintaining a fixed volume requires one cycle to increase in size while the other decreases, and so the coefficients reflecting the coupling of the associated axions to the modulus have opposite signs.

Let us now analyse the properly degenerate subspace. Taking perturbative corrections into account \cite{Berg:2005ja,vonGersdorff:2005bf,Berg:2005yu,Cicoli:2007xp}, D-brane loops lift the flat potential for $\Phi_{2}$ with a scaling
\begin{equation}
V_{\rm{loop}} \sim \frac{1}{\mathcal{V}^{10/3}},
\end{equation} 
which is parametrically faster than $\mathcal{O}(1/N)$ and thus dominates over anomalous dimensions. Therefore the degeneracy between $a_{1},a_{2}$ and $\Phi_{2}$ breaks to $a_{1}$ and $a_{2}$ only. Since there are no interactions mixing $a_1$ and $a_2$ in the Lagrangian, the resulting matrix is diagonal, with
\begin{equation}
\begin{aligned}
&\gamma(0,\ell)_{a_{1}a_{1}}\propto -f_{a_{1}a_{1}\Phi_{2}}^2 <0,\\
&\gamma(0,\ell)_{a_{2}a_{2}}\propto -f_{a_{2}a_{2}\Phi_{2}}^2 <0.\\
\end{aligned}
\end{equation}
Similarly, the anomalous dimension is negative (by construction) for $ [\Phi_{2}\Phi_{1}]_{0,\ell}$.

\subsection{Fibred Calabi Yau: Example with Two Fibred Moduli}

A similar analysis can be carried out for other fibred Calabi-Yau models (here Appendix \ref{app:2m}) gives the full details). One direct extension is to consider a model with two fibred moduli, in which the volume of the fibred Calabi-Yau manifold reads: 
\begin{equation}
\mathcal{V}=\sqrt{\tau_{1}\tau_{2}\tau_{3}}-\tau_{s}^{3/2}.
\end{equation}
Here $\tau_{1},\tau_{2},\tau_{3}$ control the volumes of individual 4-cycles, with the overall volume direction corresponding in the rescaling $\tau_{i}\rightarrow \lambda\tau_{i}$. The $\sqrt{\tau_{1}\tau_{2}\tau_{3}} $ form is reminiscent of toroidal orbifolds based on $T_{2}\times T_{2}\times T_{2}$, while $\tau_{s}$ is the blow up mode necessary to realize LVS.

In the large volume limit, the K\"ahler metric reads:
$$
K_{i\overline{j}}=\left(
\begin{matrix}
\frac{1}{4\tau_{1}^{2}} & \frac{\tau_{s}^{\frac{3}{2}}}{8\tau_{1}^{\frac{3}{2}}\tau_{2}^{\frac{3}{2}}\tau_{3}^{\frac{1}{2}}}
&\frac{\tau_{s}^{\frac{3}{2}}}{8\tau_{1}^{\frac{3}{2}}\tau_{2}^{\frac{1}{2}}\tau_{3}^{\frac{3}{2}}}&-\frac{3\tau_{s}^{\frac{1}{2}}}{8\tau_{1}\mathcal{V}} \\\\
\frac{\tau_{s}^{\frac{3}{2}}}{8\tau_{1}^{\frac{3}{2}}\tau_{2}^{\frac{3}{2}}\tau_{3}^{\frac{1}{2}}}& \frac{1}{4\tau_{2}^{2}}&
\frac{\tau_{s}^{\frac{3}{2}}}{8\tau_{1}^{\frac{1}{2}}\tau_{2}^{\frac{3}{2}}\tau_{3}^{\frac{3}{2}}}& -\frac{3\tau_{s}^{\frac{1}{2}}}{8\tau_{2}\mathcal{V}}\\\\
\frac{\tau_{s}^{\frac{3}{2}}}{8\tau_{1}^{\frac{3}{2}}\tau_{2}^{\frac{1}{2}}\tau_{3}^{\frac{3}{2}}} &\frac{\tau_{s}^{\frac{3}{2}}}{8\tau_{1}^{\frac{1}{2}}\tau_{2}^{\frac{3}{2}}\tau_{3}^{\frac{3}{2}}} & \frac{1}{4\tau_{3}^{2}}&-\frac{3\tau_{s}^{\frac{1}{2}}}{8\tau_{3}\mathcal{V}}\\\\
-\frac{3\tau_{s}^{\frac{1}{2}}}{8\tau_{1}\mathcal{V}} & -\frac{3\tau_{s}^{\frac{1}{2}}}{8\tau_{2}\mathcal{V}}& -\frac{3\tau_{s}^{\frac{1}{2}}}{8\tau_{3}\mathcal{V}} &\frac{3}{8}\frac{1}{\tau_{s}^{\frac{1}{2}}\mathcal{V}}\\\\
\end{matrix}
\right),
$$
while the inverse matrix is:
$$
K^{-1}=\left(
\begin{matrix}
4\tau_{1}^{2} & -2\frac{\tau_{s}^{\frac{3}{2}}\tau_{1}^{\frac{1}{2}}\tau_{2}^{\frac{1}{2}}}{\tau_{3}^{\frac{1}{2}}}
& -2\frac{\tau_{s}^{\frac{3}{2}}\tau_{1}^{\frac{1}{2}}\tau_{3}^{\frac{1}{2}}}{\tau_{2}^{\frac{1}{2}}}&4\tau_{1}\tau_{s} \\\\
 -2\frac{\tau_{s}^{\frac{3}{2}}\tau_{1}^{\frac{1}{2}}\tau_{2}^{\frac{1}{2}}}{\tau_{3}^{\frac{1}{2}}} &4\tau_{2}^{2}&  -2\frac{\tau_{s}^{\frac{3}{2}}\tau_{2}^{\frac{1}{2}}\tau_{3}^{\frac{1}{2}}}{\tau_{1}^{\frac{1}{2}}}& 4\tau_{2}\tau_{s}\\\\
-2\frac{\tau_{s}^{\frac{3}{2}}\tau_{1}^{\frac{1}{2}}\tau_{3}^{\frac{1}{2}}}{\tau_{2}^{\frac{3}{2}}}& -2\frac{\tau_{s}^{\frac{3}{2}}\tau_{3}^{\frac{1}{2}}\tau_{2}^{\frac{1}{2}}}{\tau_{1}^{\frac{1}{2}}} &4\tau_{3}^{2}&4\tau_{3}\tau_{s}\\\\
4\tau_{1}\tau_{s} & 4\tau_{2}\tau_{s} & 4\tau_{3}\tau_{s} & \frac{8}{3}V\tau_{s}^{\frac{1}{2}}\\\\
\end{matrix}
\right).
$$
The kinetic term for the moduli $\tau_{1},\tau_{2},\tau_{3}$ and their axion partners can be expressed as the following:
\begin{equation}
\begin{aligned}
\mathcal{L} \supset \, &\frac{1}{4\tau_{1}^{2}}\partial^{u}\tau_{1}\partial_{u}\tau_{1}+\frac{1}{4\tau_{1}^{2}}\partial^{u}a_{1}\partial_{u}a_{1}
+\frac{1}{4\tau_{2}^{2}}\partial^{u}\tau_{2}\partial_{u}\tau_{2}+\\
&\frac{1}{4\tau_{2}^{2}}\partial^{u}a_{2}\partial_{u}a_{2}+
\frac{1}{4\tau_{3}^{2}}\partial^{u}\tau_{3}\partial_{u}\tau_{3}+\frac{1}{4\tau_{3}^{2}}\partial^{u}a_{3}\partial_{u}a_{3}.\\
\end{aligned}
\end{equation}
Following similar techniques to the last section, we can substitute one of the flat moduli with the volume modulus, diagonalise the kinetic term and write down the scalar potential. After putting all the constituents together, the resulting effective field theory reads:

\begin{equation}
\begin{aligned}
\mathcal{L}= &\frac{1}{2}(\partial_{u}\Phi_{1})^{2}+\frac{1}{2}(\partial_{u}\Phi_{2})^{2}+\frac{1}{2}(\partial_{u}\Phi_{3})^{2}+\frac{1}{2}e^{-2\sqrt{\frac{2}{3}}\Phi_{1}-2\sqrt{\frac{1}{3}}\Phi_{2}-2\Phi_{3}}(\partial_{u}a_{1})^{2}\\
&+\frac{1}{2}e^{-2\sqrt{\frac{2}{3}}\Phi_{1}-2\sqrt{\frac{1}{3}}\Phi_{2}+2\Phi_{3}}(\partial_{u}a_{2})^{2}
+\frac{1}{2}e^{\left(2\sqrt{\frac{2}{3}}-2\sqrt{\frac{3}{2}}\right)\Phi_{1}+4\sqrt{\frac{1}{3}}\Phi_{2}}(\partial_{u}a_{3})^{2}-V_{LVS},\\
\end{aligned}
\end{equation}
where the axions have also been rescaled to give canonical kinetic terms. As before, the OPE coefficients can be determined with Eqs (\ref{eq:opec1})-(\ref{eq:opec2}). In particular, they are
\begin{equation}
\begin{aligned}
&f_{a_{1}a_{1}}^{\Phi_{1}}= \sqrt\frac{2}{3} \frac{\pi^{\frac{d}{2}}\Gamma\left(\frac{2\Delta_{a_{1}}+\Delta_{\Phi_{1}}-d}{2}\right)}{\Gamma\left(\Delta_{\Phi_{1}}\right)\Gamma\left(\Delta_{a_{1}}\right)^{2}}\left(\Delta_{\Phi_{1}}+2\Delta_{a_{1}}-\Delta_{a_{1}}^{2}-3\right),\\
&f_{a_{1}a_{1}}^{\Phi_{2}}=\sqrt\frac{1}{3} \frac{\pi^{\frac{d}{2}}\Gamma\left(\frac{2\Delta_{a_{1}}+\Delta_{\Phi_{2}}-d}{2}\right)}{\Gamma\left(\Delta_{\Phi_{2}}\right)\Gamma\left(\Delta_{a_{1}}\right)^{2}}\left(\Delta_{\Phi_{2}}+2\Delta_{a_{1}}-\Delta_{a_{1}}^{2}-3\right),\\
&f_{a_{1}a_{1}}^{\Phi_{3}}= \frac{\pi^{\frac{d}{2}}\Gamma\left(\frac{2\Delta_{a_{1}}+\Delta_{\Phi_{3}}-d}{2}\right)}{\Gamma\left(\Delta_{\Phi_{3}}\right)\Gamma\left(\Delta_{a_{1}}\right)^{2}}\left(\Delta_{\Phi_{3}}+2\Delta_{a_{1}}-\Delta_{a_{1}}^{2}-3\right),\\
&f_{a_{2}a_{2}}^{\Phi_{1}}=  \sqrt\frac{2}{3} \frac{\pi^{\frac{d}{2}}\Gamma\left(\frac{2\Delta_{a_{2}}+\Delta_{\Phi_{1}}-d}{2}\right)}{\Gamma\left(\Delta_{\Phi_{1}}\right)\Gamma\left(\Delta_{a_{2}}\right)^{2}}\left(\Delta_{\Phi_{1}}+2\Delta_{a_{2}}-\Delta_{a_{2}}^{2}-3\right),\\
&f_{a_{2}a_{2}}^{\Phi_{2}}= \sqrt\frac{1}{3} \frac{\pi^{\frac{d}{2}}\Gamma\left(\frac{2\Delta_{a_{2}}+\Delta_{\Phi_{2}}-d}{2}\right)}{\Gamma\left(\Delta_{\Phi_{2}}\right)\Gamma\left(\Delta_{a_{2}}\right)^{2}}\left(\Delta_{\Phi_{2}}+2\Delta_{a_{2}}-\Delta_{a_{2}}^{2}-3\right),\\
&f_{a_{2}a_{2}}^{\Phi_{3}}= -\frac{\pi^{\frac{d}{2}}\Gamma\left(\frac{2\Delta_{a_{2}}+\Delta_{\Phi_{3}}-d}{2}\right)}{\Gamma\left(\Delta_{\Phi_{3}}\right)\Gamma\left(\Delta_{a_{2}}\right)^{2}}\left(\Delta_{\Phi_{3}}+2\Delta_{a_{2}}-\Delta_{a_{2}}^{2}-3\right),\\
&f_{a_{3}a_{3}}^{\Phi_{1}}= \left( \sqrt\frac{3}{2}-\sqrt\frac{2}{3} \right) \frac{\pi^{\frac{d}{2}}\Gamma\left(\frac{2\Delta_{a_{3}}+\Delta_{\Phi_{1}}-d}{2}\right)}{\Gamma\left(\Delta_{\Phi_{1}}\right)\Gamma\left(\Delta_{a_{3}}\right)^{2}}\left(\Delta_{\Phi_{1}}+2\Delta_{a_{3}}-\Delta_{a_{3}}^{2}-3\right),\\
&f_{a_{3}a_{3}}^{\Phi_{2}}= -2\sqrt\frac{1}{3} \frac{\pi^{\frac{d}{2}}\Gamma\left(\frac{2\Delta_{a_{3}}+\Delta_{\Phi_{2}}-d}{2}\right)}{\Gamma\left(\Delta_{\Phi_{2}}\right)\Gamma\left(\Delta_{a_{3}}\right)^{2}}\left(\Delta_{\Phi_{2}}+2\Delta_{a_{3}}-\Delta_{a_{3}}^{2}-3\right),\\
&f_{a_{3}a_{3}}^{\Phi_{3}}= 0 ,\\
&f_{\Phi_{1}\Phi_{1}}^{\Phi_{1}}=\frac{81\sqrt{6}}{2}\frac{3\pi^{\frac{d}{2}}\Gamma\left(\frac{3\Delta_{\Phi_{1}}-3}{2}\right)}{\Gamma\left(\Delta_{\Phi_{1}}\right)^{3}}.
\end{aligned}
\end{equation}
Analogously to what happens in the previous examples, all the axions and the two fibre moduli are massless at zeroth order, leading to a large degeneracy between the double trace operators. A first simplification is again provided by the existence of a $\mathbb{Z}_2 \times \mathbb{Z}_2 \times \mathbb{Z}_2 $, symmetry, which partially breaks it down to subset of operators in the same representations. A classification is provided in Table \ref{tab:Z23}. 
\begin{table}
\begin{tabular}{|c|c|c|c|}
\hline  
 &$\mathbb{Z}_{2}(a_{1})$&$\mathbb{Z}_{2}(a_{2})$&$\mathbb{Z}_{2}(a_{3})$\\
\hline 
$[a_{1}a_{2}]_{n,\ell}$&-1&-1&1\\
\hline 
$[a_{1}a_{3}]_{n,\ell}$&-1&1&-1\\
\hline 
$[a_{2}a_{3}]_{n,\ell}$&1&-1&-1\\
\hline
$[a_{1}\Phi_{2}]_{n,\ell},[a_{1}\Phi_{3}]_{n,\ell}$&-1&1&1\\
\hline 
$[a_{2}\Phi_{2}]_{n,\ell},[a_{2}\Phi_{3}]_{n,\ell}$&1&-1&1\\
\hline 
$[a_{3}\Phi_{2}]_{n,\ell},[a_{3}\Phi_{3}]_{n,\ell}$&1&1&-1\\
\hline 
$[a_{1}a_{1}]_{n,\ell}, [a_{2}a_{2}]_{n,\ell}, [a_{3}a_{3}]_{n,\ell},[\Phi_{2}\Phi_{2}]_{n,\ell},[\Phi_{3}\Phi_{3}]_{n,\ell},,[\Phi_{2}\Phi_{3}]_{n,\ell}$&1&1&1\\
\hline 
\hline
$[a_{1}\Phi_{2}]_{n,\ell},[a_{1}\Phi_{1}]_{n,\ell}$&-1&1&1\\
\hline 
$[a_{2}\Phi_{2}]_{n,\ell},[a_{2}\Phi_{1}]_{n,\ell}$&1&-1&1\\
\hline 
$[a_{3}\Phi_{2}]_{n,\ell},[a_{3}\Phi_{1}]_{n,\ell}$&1&1&-1\\
\hline 
\end{tabular}
\caption{Irreducible representation of the double trace operators under the $\mathbb{Z}_{2} \times \mathbb{Z}_{2} \times \mathbb{Z}_{2} $ symmetry. The two different blocks correspond to distinct subspaces, where the degeneracies are partly broken by the discrete symmetries.}
\label{tab:Z23}
\end{table}

We are now ready to discuss the anomalous dimensions. For operators involving axions only,
\begin{equation}
\begin{aligned}
&\gamma_{a_{1}a_{2}}\propto -(f_{a_{1}a_{1}}^{\Phi_{2}}f_{a_{2}a_{2}}^{\Phi_{2}}+f_{a_{1}a_{1}}^{\Phi_{3}}f_{a_{2}a_{2}}^{\Phi_{3}})
>0,\\
&\gamma_{a_{1}a_{3}}\propto -(f_{a_{1}a_{1}}^{\Phi_{2}}f_{a_{3}a_{3}}^{\Phi_{2}})>0,\\
&\gamma_{a_{2}a_{3}}\propto -(f_{a_{2}a_{2}}^{\Phi_{2}}f_{a_{3}a_{3}}^{\Phi_{2}})>0.\\
\end{aligned}
\end{equation}
We have obtained the same, puzzling result encountered in the previous section - positive anomalous dimensions for operators involving two axions. Moreover, all three anomalous dimensions are identical in this case. On the other hand, for the ones involving the volume modulus
\begin{equation}
\begin{aligned}
&\gamma_{a_{1} \Phi_1}\propto -f_{a_{1}a_{1}}^{\Phi_{1}}f_{\Phi_{1} \Phi_{1}}^{\Phi_{1}} <0,\\
&\gamma_{a_{2} \Phi_1}\propto -f_{a_{2}a_{2}}^{\Phi_{1}}f_{\Phi_{1} \Phi_{1}}^{\Phi_{1}} <0,\\
&\gamma_{a_{3} \Phi_1}\propto -f_{a_{3}a_{3}}^{\Phi_{1}}f_{\Phi_{1} \Phi_{1}}^{\Phi_{1}} <0\\
\end{aligned}
\end{equation}
are all negative as in previous studies.
Again, all of the operators of the form $[a_{i}\Phi_{2}]_{0,\ell}, [a_{j}\Phi_{3}]_{0,\ell}$ decouple through a combination of the discrete symmetries and sub-leading volume correction, which lift the degeneracy between $\Phi_2$ and $\Phi_3$ (and with the axions too). In particular they are all zero at this order in perturbation theory. The same happens for $[\Phi_2\Phi_{2}]_{0,\ell},[\Phi_2\Phi_{3}]_{0,\ell}, [\Phi_3\Phi_{3}]_{0,\ell} $, which do not mix with the axion double trace operators for the same reason. Finally, 
$$[a_{1}a_{1}]_{n,\ell},\quad[a_{2}a_{2}]_{n,\ell},\quad[a_{3}a_{3}]_{n,\ell}$$
are still degenerate,
but the corresponding anomalous dimension matrix is diagonal and all entries are necessarily negative, since they are proportional to a sum of squared OPE coeffcients (with a minus sign in front).

\section{Type IIA Models}\label{sec:IIA}
Up to this point, we have exclusively focused on scenarios arising in type IIB compactifications: LVS, KKLT and variations thereof. However, it is clearly right that we also explore qualitatively different moduli stabilisation scenarios, in particular those arising from IIA flux compactifications. One significant difference with these is that, because IIA string theory has both 2- and 3-form flux field strengths in its spectrum, it allows for tree-level flux stabilisation of the moduli in a way that is not possible in type IIB scenarios.

In particular, we therefore consider the vacua of \cite{DeWolfe:2005uu} (for earlier work on the effective action of IIA orientifolds see \cite{Grimm:2004ua,Kachru:2004jr}). This found that type IIA Calabi-Yau orientifolds with fluxes admit stable $AdS_4$ vacua in the controlled regime of large volume and weak coupling, and  presented an explicit example where all moduli are stabilized, which will form the basis of our analysis. This construction is based on the toroidal orientifold $T^6/\mathbb{Z}^2_3$, where the discrete symmetries fix the shape of the tori and thus eliminate all complex structure moduli. Aside from the blow-up modes associated to the singularities, the only moduli are the K\"{a}hler moduli determining the size of the three tori, with their associated $B$-field axions, and the dilaton, with an axionic partner coming from the $C_3$ three-form. 

In 4D, the low energy theory is then characterized by a small number of fields which are light in the holographic sense; i.e. whose conformal dimensions remain finite in the limit of an infinite volume.  While previous investigations into the nature of the dual to this scenario \cite{Aharony:2008wz} have concentrated on more generic properties of the whole spectrum, we again focus on the detailed properties of this low-lying CFT subsector. From our perspective, the low number of moduli is thus crucial to ensure a manageable effective CFT.

\subsection{Effective Action}
In 4D, the dynamics of the type IIA vacua can be conveniently analysed with the formalism of $\mathcal{N}=1$ supergravity. The superfield coordinates corresponding to the K\"{a}hler moduli and the dilaton can be written as
\begin{equation}
t_i = b_i + i v_i \quad \quad S = e^{-D}+ i \frac{\xi}{\sqrt{2}},
\end{equation}
and they are related to the overall volume of the manifold $\mathcal{V}$ and the 10-dimensional dilaton $\phi$ by
\begin{equation}\label{eq:vmin}
\mathcal{V}= \kappa \, v_1 v_2 v_3,\quad \quad \quad e^D = \frac{e^{\phi}}{\sqrt{\mathcal{V}}}.
\end{equation}
 In terms of these variables, the K\"{a}hler potential is \cite{Grimm:2004ua,DeWolfe:2005uu}
\begin{equation}
\mathcal{K}= -\log(\mathcal{V})- 4\log(S+\bar{S}).
\end{equation}
This gives rise to a kinetic term
\begin{equation}\label{eq:kt}
\mathcal{L} \supset   \partial_{\mu} D \partial^{\mu}  D +\frac{1}{2}e^{2D} \partial_{\mu} \xi \partial^{\mu}  \xi +\sum_i \frac{1}{2} \partial_{\mu} \phi_i \partial^{\mu}  \phi_i+ \frac{1}{4}e^{-2 \sqrt{2} \phi_i} \partial_{\mu} b_i \partial^{\mu}  b_i,
\end{equation}
where the canonically normalized fields $\phi_i = \frac{1}{\sqrt{2}}\log (v_i)$ have been used to make contact with the conventions of \cite{DeWolfe:2005uu}.\footnote{For the same reason the 4-dimensional dilaton has not yet been normalized.}
The moduli are then stabilized by turning on background fluxes for the NS-NS 3-form $H_3$, a constant $F_0$ and the four-form $F_4$,  quantized according to
\begin{equation}
\int F_{p}= (2 \pi)^{p-1} \alpha^{\prime(p-1) / 2} f_{p}.
\end{equation}
These three fluxes can also be rewritten in terms of integers $f_p$ as
\begin{equation}
m_{0}=\frac{f_{0}}{2 \sqrt{2} \pi \sqrt{\alpha^{\prime}}}, \quad p=(2 \pi)^{2} \alpha^{\prime} h_{3}, \quad e_{i}=\frac{\kappa^{1 / 3}}{\sqrt{2}}\left(2 \pi \sqrt{\alpha^{\prime}}\right)^{3} f_{4}^{i},
\end{equation}
where the index $i$ in the last equation refers to the three non-trivial 4-cycles of the manifold. To ensure tadpole cancellation, the first two fluxes cannot be independent and must satisfy
\begin{equation}
m_{0} p= -2\left(\sqrt{2} \pi \sqrt{\alpha^{\prime}}\right).
\end{equation}
In terms of these quantities, one can finally write a scalar potential for the moduli \cite{DeWolfe:2005uu}, which reads
\begin{equation}
V= \frac{p^2}{4} \frac{e^{2D}}{k} e^{-\sqrt{2} \sum_i \phi_i } +  \Big( \sum_i e_i^2 e^{2\sqrt{2} \phi_i} \Big) \frac{e^{4D-\sqrt{2}\sum_i \phi_i}}{2k }+\frac{m_0^2}{2} e^{4D}k e^{\sqrt{2}\sum_i \phi_i}-\sqrt{2} |m_0 p| e^D.
\end{equation}
In particular, the minimum occurs for
\begin{equation}\label{eq:min1}
\bar{\phi}_i = \frac{1}{\sqrt{2} } \log \Bigg( \frac{1}{|e_i|}\sqrt{\frac{5}{3} \abs*{\frac{e_1 e_2 e_3}{k m_0}}}\,\Bigg )
\end{equation} 
\begin{equation}\label{eq:min2}
e^D = |p| \sqrt{\frac{27}{160}\abs*{\frac{k m_0}{e_1 e_2 e_3}}},
\end{equation}
where the potential takes the value
\begin{equation}
V|_{\text{min}}\equiv - \frac{3}{R^2_{AdS}} = - \frac{243 \sqrt{\frac{3}{5}} k^{3/2} p^4 \left| m_0\right| {}^{5/2}}{12800 \abs{e_1 e_2 e_3}^{3/2}}.
\end{equation}
In the presence of background fluxes, dimensional reduction of the $B_2$ and $C_3$ forms gives rise to a potential for the axions. The effective Lagrangian of the form \cite{DeWolfe:2005uu}
\begin{equation}\label{eq:axlag}
\begin{split}
\mathcal{L}\ \supset\,   &\,  \frac{1}{4} \sum_{i=1}^{3}  e^{-2 \sqrt{2}\phi_i}  \partial_{\mu} b_i \partial^{\mu} b_i +
\frac{1}{2} e^{2D} \partial_{\mu} \xi \partial^{\mu} \xi -\frac{e^{4 D}}{\mathcal{V}}\left(b_{1} e_{1} +b_{2} e_{2} +b_{3} -p \xi \right)^{2}  \\
& -\frac{e^{4 D}}{2} \sum_{i=1}^{3} \Big( m_{0}^{2}  e^{-2  \sqrt{2}\phi_i} \mathcal{V} \, b_{i}^{2} -2 m_{0}e^{2\sqrt{2}\phi_i-\sqrt{2}(\phi_1+\phi_2+\phi_3)} b_{1} b_{2} b_{3} \frac{e_{i} }{b_{i}}  \Big),
\end{split}
\end{equation}
\subsection{Integer Conformal Dimensions}
Expanding around the vacuum solution discussed in the previous section, we can obtain the masses and couplings for the moduli. In particular, the mass matrix is non diagonal; full details of the mixing are provided in Appendix \ref{sec:A}. Here, we stress a surprising fact: after diagonalization, the dilatation operator eigenstates obtain conformal dimensions which do not depend on any of the compactification parameters (e.g. the values of the fluxes) and furthermore are all integers:
\begin{equation}
\Delta_{\varphi}= (10,6,6,6).
\end{equation} 

Similarly, the couplings appearing in the interaction terms are fixed only by the AdS radius, and all come with the same factor of $R_{AdS}^{-2}$ in front. 
As in LVS, this again hints at the fact that the dual CFT at large volumes can be thought as small perturbation above some universal generalized free theory. There is also a high degree of degeneracy between the various modes. Naively this would appear to translate into mixing between the corresponding double trace operators at order $\mathcal{O}(1/N^2)$.\footnote{Notice that is also true for the first scalar, since the spectra of different double trace operators overlap when their conformal dimensions are integer spaced (and not only when they are equal). } However, in practice we expect this degeneracy to be lifted by the effects of higher order operators. The degeneracy partially comes as a result of the exchange symmetries between pairs of tori and their fluxes $b_i \rightarrow b_j$ and $e_i \rightarrow e_j$\footnote{And the independence from the fluxes, which effectively turns the symmetry into $b_i \rightarrow b_j$ only.}, but the fact that there is also a third identical eigenvalue seems to be coincidental. 

Surprisingly, the exact same phenomenon occurs for the K\"ahler axions, with the only difference that there is a residual imprint of the flux signs, and one can distinguish between a few discrete cases. Expanding (\ref{eq:axlag}) around the solution (\ref{eq:min1}) and (\ref{eq:min2}), the mass matrix is almost independent of the values taken by the fluxes, which only enter through the signs
\begin{equation}
s_i \equiv \text{sgn}(m_0 e_i).
\end{equation}
While different choices of signs do not affect the vacuum, they do affect its supersymmetry - the only choice that leads to a supersymmetric vacuum is complete negativity, $s= (-1,-1,-1)$ (e.g. see \cite{Narayan:2010em}). However, for our purposes the presence of supersymmetry is not crucial either way.

The diagonalisation procedure is reported in Appendix \ref{sec:A}, where we also give the full set of cubic interactions relevant for the computation of anomalous dimensions. Except for a discrete sign choice associated to the $s_i$'s, couplings are entirely determined up to a global $R_{AdS}^{-2}$ factor. For the conformal dimensions, the signs $s= (1,1,1)$ and $s= (1,-1,-1)$ yield
\begin{equation}
\Delta_a = (8,8,8,2)  \quad\quad \text{or} \quad\quad \Delta_a = (8,8,8,1).
\end{equation}
Two choices are available for the last eigenvalue, since both solutions for the conformal dimension in terms of the mass satisfy the unitarity bound $\Delta \geq d/2-1$. For the other two cases $s= (-1,1,1)$ and $s= (-1,-1,-1)$, there is only one possibility:
\begin{equation}
\Delta_a = (11,5,5,5).
\end{equation} 
In both instances, the considerations about degeneracy apply as before, and again we only see integer conformal dimensions. 

While we cannot offer an explanation of this fact within our framework, it is certainly surprising to see the emergence of such a peculiar pattern. Given the quadratic nature of the $\Delta (\Delta-d) = m^2 R_{AdS}^2$ relation, the origin of these integers cannot simply be traced back to integers on the AdS, giving more credibility to the idea that the CFT duals might offer a complementary perspective on the description of such vacua. Moreover, the (almost) complete independence on the details of the compactification is already a striking feature, also shared with LVS (but not KKLT). It is very much in line with the philosophy of the Swampland program, according to which consistency conditions should become more and more stringent in certain limits, and eventually culminate with a very restricted set of ``special'' theories.

Interestingly, something similar happens if we consider what happens for more general Calabi-Yaus with $h^{2,1}$ complex structure moduli $U_i$. The above example concerned a $\mathbb{T}^6 / \mathbb{Z}_3 \times \mathbb{Z}_3$ geometry with no complex structure moduli at all. For more general IIA flux vacua, an orientifold projection reduces the original hypermultiplet down to two scalars. The real part corresponds to the calibrated volume of a 3-cycle $\Sigma_i$ weighted by $g_s^{-1}$, and the imaginary part is a `true' axion arising from the reduction of the RR 3-form on the cycle, $\int_{\Sigma_i} C_3$. For models of particle physics arising from intersecting D6-branes, these $U_i$ moduli represent the gauge kinetic functions for the gauge groups living on the D6 branes.
$$
f_a = \sum \lambda_i U_i.
$$
The use of the expression `true' axion refers to the origin from an RR 3-form. Within weakly coupled string theory, the only physics sensitive to the value of such axions are non-perturbative effects - either D-brane instantons or gaugino condensation - associated to the cycles wrapped by the D6 branes. All such effects are suppressed by non-perturbative exponentials, $e^{-2 \pi \textrm{Vol}(\Sigma_i)/g_s }$, which are non-perturbative in both $\alpha'$ and $g_s$ expansions. This implies that, within perturbation theory, the shift symmetry $\textrm{Im}(U_i) \to \textrm{Im}(U_i) + \epsilon$ is exact: if it is there at tree-level, it is also there at all orders in perturbation theory.

In IIA flux compactifications, the superpotential for the complex structure moduli is linear in the complex structure moduli. Equivalently, the presence of the fluxes \emph{explicitly} breaks the shift symmetry for one -- and only one -- linear combination of the axions. By taking linear combinations of the moduli, one can regard the superpotential as depending on only one of these moduli, while the remaining 
moduli do not appear in the superpotential and also retain the shift symmetry, $K = K(U_i + \bar{U}_i)$.

For such moduli, their axionic components are of course massless (as nothing in the potential depends on their vev). However, at the minimum the properties of the saxions are also fixed. In fact, one can show that for a supersymmetric minimum, the above properties imply (\cite{Conlon:2006tq} for a general argument, \cite{Marchesano:2019hfb,Marchesano:2020uqz} for explicit spectra and some non-supersymmetric examples)
$$
M_{saxion}^2 = - \frac{2}{3} V_{min} \equiv \frac{-2}{R_{AdS}^2}.
$$
On a holographic interpretation, this immediately implies
$$
\Delta_{saxion} \left( \Delta_{saxion} - 3 \right) = -2
$$
and so
$$
\Delta_{saxion} = 1 \qquad \textrm{   or  } \qquad \Delta_{saxion} = 2.
$$

\subsection{Degeneracy Lifting and Anomalous Dimensions}
As explained in \ref{ssc:deg}, the presence of degeneracies can significantly complicate the calculation of anomalous dimensions, because of the mixing between the double trace operators occurring at $\mathcal{O}(1/N^2)$. However, the effect is not present if the conformal dimensions of the single trace operators receive corrections of order $\mathcal{O}(1/N)$ or higher, as this effectively lifts the degeneracy. This is what we would expect to happen in this case, as can be inferred from the scaling of higher derivative supergravity corrections. From Eq. (\ref{eq:min1}), the controlled limit of large volume is obtained for large four-form fluxes, when $\abs{e_i} >> \abs{m_0}$. Assuming all fluxes to be of the same order $e_i \sim \bar{e}$, we then have the following scalings:
\begin{equation}
v_i \sim \bar{e}^{\frac{1}{2}} \quad \quad  e^D \sim \bar{e}^{-\frac{3}{2}}\quad \quad R_{AdS} \sim \bar{e}^{\frac{9}{4}} \quad \quad \frac{1}{N} \sim \bar{e}^{-\frac{9}{2}},
\end{equation}
which can be used to estimate the impact of higher dimensional operators.
As noted in \cite{Grimm:2004ua}, the scalar potential will receive relative corrections of order $\bar{e}^{-\frac{3}{2}}$ from $\abs{F_4}^4$ terms in the Lagrangian, taking all the metric contractions and $g_s$ factors into account. This will result in a mass shift for the moduli 
\begin{equation}
\delta \Delta_{\varphi} =\frac{2 \Delta_{\varphi} (\Delta_{\varphi}-3)}{2 \Delta_{\varphi}-3} \bigg( \frac{\delta m^2}{m^2} \bigg) \sim \bar {e}^{-\frac{3}{2}} \gg \frac{1}{N}.
\end{equation}
Similarly, $\abs{F_4}^2 \abs{B_2}^2$ terms will be multiplied by additional factors of $g_s$ and $R^{-4}$, and give
\begin{equation}
\delta \Delta_a =\frac{2 \Delta_a(\Delta_a-3)}{2 \Delta_a-3} \bigg( \frac{\delta m^2}{m^2} \bigg) \sim \bar {e}^{-\frac{5}{2}} \gg \frac{1}{N}.
\end{equation}
for the axions. This would suggest that the degeneracy is lifted in both cases, and anomalous dimensions can be calculated using the standard procedure.\footnote{We should give one note of caution here. As we do not understand the origin of integer conformal dimensions, we cannot exclude the possibility that whatever structure gave rise to them also holds for the higher order terms - in which case the higher order terms would fail to lift the degeneracy.}

Unfortunately, it also means that the single trace operators, out of which the double trace ones are constructed, will be rotated with respect to the basis we are using. Within an exactly degenerate subspace (with two or more identical conformal dimensions), the eigenstates will depend on the degeneracy-splitting hamiltonian at leading order, and it is impossible to deduce what they are without knowing the explicit form of the perturbation. Thus, the only meaningful anomalous dimensions correspond to the double trace operators constructed with singlets of the original Hamiltionian: $\varphi_1$ and $a_1$ or $a_4$ (corresponding to cases 1 and 2, which depend on the flux signs). As shown in appendix \ref{sec:A}, cases 3 and 4 can be related to and 2 respectively by some discrete transformations which leave the anomalous dimensions invariant, so we can focus only on the first two. In the first case, formula (\ref{eq:adim}) dictates that the anomalous dimension will be proportional to a factor
\begin{equation}
\gamma_{\varphi_1 a_1 } \propto \frac{1}{ \Gamma(\Delta_{a_1}-\frac{\Delta_{\mathcal{O}}}{2})   },
\end{equation} 
where $\mathcal{O}$ is the lowest dimensional operator exchanged (one of $\varphi_2,\varphi_3,\varphi_4$ in this case, with $\Delta_{\varphi_i}=6$). For the values of $\Delta$ found above this term diverges, and it is only regulated by sub-leading corrections to the conformal dimensions. However, given the behaviour of the gamma function near its poles, the sign of the anomalous dimension will depend on the signs of such corrections, and cannot be determined at this point. The only thing that can be said is that the two cases $\Delta_a = 1$ and $\Delta_a = 2$ the overall sign will be the same, since any correction will shift their magnitude in opposite directions. On the other hand, the relevant anomalous dimension in case 2 can be calculated meaningfully. Using the couplings derived in \ref{sec:A}, 
\begin{equation}
C_{\varphi_1 \varphi_1 \varphi_2 } = -\frac{315 \sqrt{2}}{169}  \quad \quad 
C_{\varphi_1 \varphi_1 \varphi_3 } = \frac{105 \sqrt{2}}{169} \quad \quad 
C_{\varphi_1 \varphi_1 \varphi_4 } = \frac{105 \sqrt{2}}{169}  \quad \quad 
\end{equation}
and 
\begin{equation}
C_{a_1 a_1 \varphi_2 } = -\frac{60(3+100\sqrt{2})}{2197}\quad \,\, 
C_{a_1 a_1 \varphi_3 } = \frac{20(3+100\sqrt{2})}{2197}\quad \,\,
C_{a_1 a_1 \varphi_4 } =\frac{20(3+100\sqrt{2})}{2197}  \quad \,\, .
\end{equation}
Since the kinematic factors appearing in (\ref{eq:adim}) are always positive in this example,
\begin{equation}
\gamma_{\varphi_1 a_1}(0,\ell) \propto - \sum_{i=2}^4  C_{\varphi_1 \varphi_1 \varphi_i } C_{a_1 a_1 \varphi_i } < 0
\end{equation}
Thus the only anomalous dimensions we can calculate, which involve an axion and a modulus (as in LVS), are negative. 

\section{Conclusions}\label{sec:Conclusions}

We summarise here what we regard as the most interesting results and open questions from these investigations.

First, in the context of the type IIA models of flux stabilisation the conformal dimensions of low-lying operators dual to moduli fields take on a form that is both universal and surprising, in that in a large-volume limits the dimensions are all integer and (modulo some discrete flux choices) fixed. As well as the conformal dimensions, it is also the case that the higher-point interactions take on a universal form in the large-volume limit and `forget' about the precise details of flux quantum numbers.

There are two particular aspects of this that are striking. First, the universality of the large-volume limit is reminiscent of similar behaviour for LVS. There, although the dual moduli dimensions were non-integer, the CFT also took a universal form in the large-volume limit. It is worth stressing that this runs counter to the landscape picture of string vacua being able to scan over enormously large parameter spaces in the low-energy effective field theory. Instead, these two examples of moduli stabilisation scenarios producing vacua that are in the extreme limit of K\"ahler moduli space instead lead to almost unique forms for the corresponding dual CFT.

We note this is \emph{not} true of KKLT or racetrack stabilisation, which do not stabilise in the asymptotic regions of moduli space.
Even with extraordinarily small values of $W_0$ (such as the $10^{-95}$ reported in \cite{Demirtas:2021nlu}) the volume in KKLT is only logarithmic in $W_0$, and so is never in a limit of asymptotically large volume.
 
It is intriguing to speculate that any stabilisation method producing vacua in the asymptotic large-volume limit (or, more generally, in the extreme limits of moduli space) corresponds to a very limited set of choices for the CFT properties. This sounds vaguely similar to ideas for classifying the extremal limits of moduli space (e.g. see \cite{Grimm:2019ixq}) and a more precise correspondence would be interesting.

The other striking aspect is the presence of integer dual conformal dimensions for the moduli in IIA flux vacua. This is surprising, as we can discern no obvious reason why this should be the case. In information, the conformal dimensions are equivalent to the moduli masses, but it is only when expressed in this particular fashion that an integer structure appears - the masses are not, for example, integer multiples of $R_{AdS}^{-1}$ or $m_{BF}$. This hints at some deeper structure, but we can offer no explanation as to what it is.

A second interesting result concerns the sign of the anomalous dimensions. In our previous work, we found that all anomalous dimensions for the double-trace operators in the moduli sector were negative, and this negativity correlated to signs in the low-energy effective field theory that were required from swampland considerations (for example, a sign change would correspond to a divergent axion decay constant in the infinite volume limit). However, in this case (in the IIB fibred construction), this is only true for operators involving an axion and a modulus. Indeed, all the double trace operators made out of two separate axions now have positive anomalous dimensions.

\section*{Acknowledgements}

We thank Sandipan Kundu and especially Pietro Ferrero for illuminating conversations and STFC for financial support of the Oxford particle theory group. SN acknowledges funding support from the China Scholarship Council-FaZheng Group- University of Oxford. FR is supported by the Dalitz Graduate Scholarship, jointly established by the Oxford University Department of Physics and Wadham College.

\appendix
\section{Anomalous Dimension Matrix for the Degenerate Case}\label{appendix:deg}
In this appendix we show how the anomalous dimensions of a degenerate system can be obtained by diagonalizing the matrix of the ``effective" anomalous dimensions $ \tilde{\gamma}_{[ij]\,[kl]}(n,\ell)$, as stated in section
\ref{ssc:ofpb}. The latter are obtained from the mixed correlators $\langle \mathcal{O}_i\mathcal{O}_j\mathcal{O}_k\mathcal{O}_l \rangle$, and will be defined precisely in the following.

From a bootstrap perspective, the crucial point is that at zeroth order in the large-$N$ expansion the contribution of the various double traces in any of the OPEs cannot be distiguished from one another. Schematically, the $\mathcal{O}_1\mathcal{O}_1, \, \mathcal{O}_2\mathcal{O}_2$ and $\, \mathcal{O}_1\mathcal{O}_2$ OPEs' can be decomposed as
\begin{equation}\label{eq:ope}
\mathcal{O}_i(x) \mathcal{O}_j(0) \supset  c_{ijA}(n,\ell)\,[\mathcal{O} \mathcal{O}]^A_{n,\ell}(x); \,\, c_{ijB}(n,\ell)\,[\mathcal{O} \mathcal{O}]^B_{n,\ell}(x); \,\, c_{ijC}(n,\ell)\,[\mathcal{O} \mathcal{O}]^C_{n,\ell}(x) 
\end{equation}
where the $[\mathcal{O} \mathcal{O}]^{A,B,C}_{n,\ell}$ operators are orthogonal linear combinations of the double trace operators in absence of mixing, which diagonalize the Hamiltonian at order $1/N^2$.

In terms of these variables, the conformal block expansion of a correlator involving pairs of identical operators is
\begin{equation}
\langle \mathcal{O}_i\mathcal{O}_j\mathcal{O}_i\mathcal{O}_j \rangle = \sum_{\Delta, \ell} c_{ijA}^2(n,\ell)G^{A}_{\Delta,\ell} + c_{ijB}^2(n,\ell)G^{B}_{\Delta,\ell} + c_{ijC}^2(n,\ell)G^{C}_{\Delta,\ell},
\end{equation}
where the $G^{X}_{\Delta,\ell}$ are the conformal blocks corresponding to the $[\mathcal{O} \mathcal{O}]^X_{n,\ell}(x)$ double trace operators. The functional form of the conformal blocks is identical in all cases, and the subscript is just to remind they are a function of different conformal dimensions. Up to order $\mathcal{O} \big (\frac{1}{N^4} \big)$, each term can be expanded as:
\begin{equation}
\begin{split}
& \big[ c_{ijA}^{(0)\, 2}(n,\ell)+c_{ijB}^{(0)\, 2}(n,\ell)+ c_{ijC}^{(0)\, 2}(n,\ell) \big] G_{2\Delta+2n+\ell,\ell}   \\
+&  \big[2  c^{(0)}_{ijA}  c^{(1)}_{ijA} +  2c^{(0)}_{ijB}  c^{(1)}_{ijB}  +2 c^{(0)}_{ijC}  c^{(1)}_{ijC}  \big] \frac{G_{2\Delta+2n+\ell,\ell}}{N^2} \\
 + &  \big[  c_{ijA}^{(0)\, 2}\gamma_A(n,\ell)+  c_{ijB}^{(0)\, 2}(n,\ell)\gamma_B(n,\ell)+ c_{ijC}^{(0)\, 2}\gamma_C(n,\ell) \big] \frac{\partial_{\Delta} G_{2\Delta+2n+\ell,\ell}}{N^2}  \\
\end{split}
\end{equation}
Comparing this with the standard expansion for the correlator 
\begin{equation}
 f^{(0)\,2}(n,\ell) G^{11}_{2\Delta+2n+\ell,\ell} +2 f^{(0)}(n,\ell) f^{(1)} (n,\ell) \frac{G_{2\Delta+2n+\ell,\ell}}{N^2} +  f^{(0)\,2}(n,\ell)\tilde{\gamma}_{[ij][ij]}(n,\ell)
 \frac{\partial_{\Delta} G_{2\Delta+2n+\ell,\ell}}{N^2} ,
\end{equation}
we obtain the equations
\begin{align}
\label{eq:ope1}
& c_{ijA}^{(0)\, 2}(n,\ell)+c_{ijB}^{(0)\, 2}(n,\ell)+ c_{ijC}^{(0)\, 2}(n,\ell) = f^{(0)\,2}(n,\ell)\\
\label{eq:adm1}
& c_{ijA}^{(0)\, 2}\gamma_A(n,\ell)+  c_{ijB}^{(0)\, 2}(n,\ell)\gamma_B(n,\ell)+ c_{ijC}^{(0)\, 2}\gamma_C(n,\ell) = f^{(0)\,2}(n,\ell) \tilde{\gamma}_{[ij][ij]}(n,\ell)
\end{align}
The same can be done for a mixed correlator $ \langle \mathcal{O}_i\mathcal{O}_j \mathcal{O}_k \mathcal{O}_l \rangle$, with different pairs $(i,j)\neq (k,l)$. Then, the conformal block decomposition reads 
\begin{equation}
\langle \mathcal{O}_i\mathcal{O}_j\mathcal{O}_i\mathcal{O}_j \rangle = \sum_{\Delta, \ell} c_{ijA}(n,\ell)c_{klA}(n,\ell)G^{A}_{\Delta,\ell} + c_{ijB}(n,\ell)c_{klB}(n,\ell)G^{B}_{\Delta,\ell} + c_{ijC}c_{klC}(n,\ell)G^{C}_{\Delta,\ell}.
\end{equation}
and can be expanded as 
\begin{equation}
\begin{split}
& \big[ c_{ijA}^{(0)}(n,\ell)c_{klA}^{(0)}(n,\ell)+c_{ijB}^{(0)}(n,\ell)c_{klB}^{(0)}(n,\ell)+ c_{ijC}^{(0)}(n,\ell)c_{klC}^{(0)}(n,\ell) \big] G_{2\Delta+2n+\ell,\ell}   \\
+&  \big[  c^{(0)}_{ijA}  c^{(1)}_{klA} +c^{(1)}_{ijA}  c^{(0)}_{klA} + c^{(0)}_{ijB}  c^{(1)}_{klB} +c^{(1)}_{ijB}  c^{(0)}_{klB} +c^{(0)}_{ijC}  c^{(1)}_{klC} +c^{(1)}_{ijC}  c^{(0)}_{klC} \big] \frac{G_{2\Delta+2n+\ell,\ell}}{N^2} \\
 + &  \big[  c_{ijA}^{(0)} c_{klA}^{(0)}\gamma_A(n,\ell)+  c_{ijB}^{(0)} c_{klB}^{(0)}\gamma_B(n,\ell)+ c_{ijC}^{(0)} c_{klC}^{(0)}\gamma_C(n,\ell) \big] \frac{\partial_{\Delta} G_{2\Delta+2n+\ell,\ell}}{N^2}.  \\
\end{split}
\end{equation}
Since the disconnected four-point function only receives a contribution from the identity, the same correlator can be written as 
\begin{equation}
 2 f^{(0)}(n,\ell) f^{(1)} (n,\ell) \frac{G_{2\Delta+2n+\ell,\ell}}{N^2} +  f^{(0)\,2}(n,\ell)\tilde{\gamma}_{[ij][kl]}(n,\ell)
 \frac{\partial_{\Delta} G_{2\Delta+2n+\ell,\ell}}{N^2},
\end{equation}
resulting in the equations
\begin{align}
\label{eq:ope2}
& c_{ijA}^{(0)}(n,\ell)c_{klA}^{(0)}(n,\ell)+c_{ijB}^{(0)}(n,\ell)c_{klB}^{(0)}(n,\ell)+ c_{ijC}^{(0)}(n,\ell)c_{klC}^{(0)}(n,\ell) = 0\\
\label{eq:adm2}
& c_{ijA}^{(0)}(n,\ell)c_{klA}^{(0)}(n,\ell) \gamma_A(n,\ell)+  c_{ijB}^{(0)\, 2}(n,\ell)\gamma_B(n,\ell)+ c_{ijC}^{(0)\, 2}\gamma_C(n,\ell) = f^{(0)\,2}(n,\ell) \tilde{\gamma}_{[ij][ij]}(n,\ell).
\end{align}
This system admits a simple geometric interpretation. Equations (\ref{eq:ope1}) and (\ref{eq:ope2}) imply that the vectors 
\begin{equation}
\vec{c}_{ij}^{(0)}(n,\ell) = \frac{1}{f^{(0)}(n,\ell)} \big(c_{ijA}^{(0)}(n,\ell),c_{ijB}^{(0)}(n,\ell),c_{ijC}^{(0)}(n,\ell) \big)
\end{equation}
are an orthonormal basis in $\mathbb{R}^3$. In this basis, the original anomalous dimension matrix $\mathbf{M} (n,\ell)= \rm{Diag}(\gamma_A(n,\ell),\gamma_B(n,\ell),\gamma_C(n,\ell))$ has matrix elements
\begin{equation}
\mathbf{\tilde{M}}(n,\ell)_{[ij][kl]} = \bra{c_{ij}^{(0)}(n,\ell)} \mathbf{M} (n,\ell) \ket{c_{kl}^{(0)}(n,\ell)} = \tilde{\gamma}_{[ij][kl]},
\end{equation}
where the second equality derives from (\ref{eq:adm1}) and  (\ref{eq:adm2}). Therefore, the $\tilde{\gamma}_{[ij][kl]}$ are the coefficients of the anomalous dimension matrix $\mathbf{M} (n,\ell)$ expressed in a different orthonormal basis, implying the two share the same eigenvalues.

\section{Fibred Calabi-Yau Effective Lagrangians}
\subsection{One Fibre Modulus}\label{app:1m}
From equation(\ref{eq1}), the Lagrangian for the fibred one-modulus Calabi Yau scenario is :
\begin{equation}
L_{kin}=\frac{3}{8\tau_{1}^{2}}\partial_{u}\tau_{1}\partial^{u}\tau_{1}-\frac{1}{2\tau_{1}\mathcal{V}}\partial_{u}\tau_{1}\partial^{u}\mathcal{V}+\frac{1}{2\mathcal{V}^{2}}\partial_{u}\mathcal{V}\partial^{u}\mathcal{V}
+\frac{1}{4\tau_{1}^{2}}\partial_{u}a_{1}\partial^{u}a_{1}+\frac{\alpha^{2}\tau_{1}}{2\mathcal{V}^{2}}\partial_{u}a_{2}\partial^{u}a_{2}.
\end{equation}
To put this in a canonically normalised form, we use the ansatz:
\begin{equation}
\begin{aligned}
&\textrm{ln}\tau_{1}=a\Phi_{1}+b\Phi_{2},\\
&\textrm{ln}\mathcal{V}=c\Phi_{1}.\\
\end{aligned}
\end{equation}
We subsitute this ansatz into the Lagrangian:
\begin{equation}
\begin{aligned}
\frac{3}{8\tau_{1}^{2}}\partial_{u}\tau_{1}\partial^{u}\tau_{1}&=\frac{3}{8}(\partial_{u}\textrm{ln}\tau_{1})^{2}
=\frac{3}{8}(a\partial_{u}\Phi_{1}+b\partial_{u}\Phi_{2})^{2}\\
&=\frac{3}{8}a^{2}(\partial_{u}\Phi_{1})^{2}+\frac{3}{8}b^{2}(\partial_{u}\Phi_{2})^{2}+\frac{3}{4}ab\partial_{u}\Phi_{1}\partial^{u}\Phi_{2},\\
\end{aligned}
\end{equation}
\begin{equation}
\begin{aligned}
-\frac{1}{2\tau_{1}\mathcal{V}}\partial_{u}\tau_{1}\partial^{u}\mathcal{V}&=
-\frac{1}{2}\partial_{u}\textrm{ln}\tau_{1}\partial^{u}\textrm{ln}\mathcal{V}
=-\frac{1}{2}(a\partial_{u}\Phi_{1}+b\partial_{u}\Phi_{2})c\partial^{u}\Phi_{1}\\
&=-\frac{1}{2}ac(\partial_{u}\Phi_{1})^{2}-\frac{1}{2}bc\partial_{u}\Phi_{2}\partial^{u}\Phi_{1},\\
\end{aligned}
\end{equation}
\begin{equation}
\begin{aligned}
\frac{1}{2\mathcal{V}^{2}}\partial_{u}\mathcal{V}\partial^{u}\mathcal{V}=
\frac{1}{2}(\partial_{u}\textrm{ln}\mathcal{V})^{2}
=\frac{1}{2}c^{2}(\partial_{u}\Phi_{1})^{2}.
\end{aligned}
\end{equation}
Putting them together, the kinetic term reads:
\begin{equation}
L_{kin}=\left(\frac{3}{8}a^{2}-\frac{1}{2}ac+\frac{1}{2}c^{2}\right)(\partial_{u}\Phi_{1})^{2}
+\left(\frac{3}{4}ab-\frac{1}{2}bc\right)\partial_{u}\Phi_{1}\partial^{u}\Phi_{2}
+\frac{3}{8}b^{2}(\partial_{u}\Phi_{2})^{2}
\end{equation}
Canonical normalisation then requires satisfying the following conditions:
\begin{equation}
\frac{3}{8}a^{2}-\frac{1}{2}ac+\frac{1}{2}c^{2}=\frac{1}{2},
\end{equation}
\begin{equation}
\frac{3}{4}ab-\frac{1}{2}bc=0,
\end{equation}
\begin{equation}
\frac{3}{8}b^{2}=\frac{1}{2}.
\end{equation}
A solution is:
\begin{equation}
\begin{aligned}
&a=\sqrt{\frac{2}{3}},\\
&b=\frac{2}{\sqrt{3}},\\
&c=\sqrt{\frac{3}{2}}.\\
\end{aligned}
\end{equation}
After substituting this into the kinetic term, we have:
\begin{equation}
\frac{3}{8\tau_{1}^{2}}\partial_{u}\tau_{1}\partial^{u}\tau_{1}
=\frac{1}{4}(\partial_{u}\Phi_{1})^{2}+\frac{1}{2}(\partial_{u}\Phi_{2})^{2}+\frac{\sqrt{2}}{2}\partial_{u}\Phi_{1}\partial^{u}\Phi_{2},
\end{equation}
\begin{equation}
-\frac{1}{2\tau_{1}\mathcal{V}}\partial_{u}\tau_{1}\partial^{u}\mathcal{V}
=-\frac{1}{2}(\partial_{u}\Phi_{1})^{2}-\frac{\sqrt{2}}{2}\partial_{u}\Phi_{1}\partial^{u}\Phi_{2},
\end{equation}
\begin{equation}
\frac{1}{2\mathcal{V}^{2}}\partial_{u}\mathcal{V}\partial^{u}\mathcal{V}
=\frac{1}{2}(\partial_{u}\ln \mathcal{V})^{2}
=\frac{3}{4}(\partial_{u}\Phi_{1})^{2}.
\end{equation}
Adding these together, we have both diagonal kinetic terms and diagonal mass terms
\begin{equation}
\frac{1}{2}(\partial_{u}\Phi_{1})^{2}+\frac{1}{2}(\partial_{u}\Phi_{2})^{2}+V(\Phi).
\end{equation}

Where the canonical  fields $\Phi_{1}, \Phi_{2}$ relate to the original fields as:
\begin{equation}
\mathcal{V}=e^{\sqrt{\frac{3}{2}}\Phi_{1}},
\end{equation}
\begin{equation}
\tau_{1}=e^{\sqrt{\frac{2}{3}}(\Phi_{1}+2\Phi_{2})}.
\end{equation}

\subsection{Two Fibre Moduli}\label{app:2m}
We again start from the kinetic terms in the Lagrangian:
\begin{equation}
\begin{aligned}
L_{kin}=&\frac{1}{4\tau_{1}^{2}}\partial^{u}\tau_{1}\partial_{u}\tau_{1}+\frac{1}{4\tau_{1}^{2}}\partial^{u}a_{1}\partial_{u}a_{1}
+\frac{1}{4\tau_{2}^{2}}\partial^{u}\tau_{2}\partial_{u}\tau_{2}+\\
&\frac{1}{4\tau_{2}^{2}}\partial^{u}a_{2}\partial_{u}a_{2}+
\frac{1}{4\tau_{3}^{2}}\partial^{u}\tau_{3}\partial_{u}\tau_{3}+\frac{1}{4\tau_{3}^{2}}\partial^{u}a_{3}\partial_{u}a_{3}\\
\end{aligned}
\end{equation}

To keep the mass term diagonal, we must have:
\begin{equation}
\ln \mathcal{V}=a\Phi_{1},
\end{equation}
and we also anticipate:
\begin{equation}
\begin{aligned}
&\textrm{ln}\tau_{1}=b\Phi_{1}+c\Phi_{2}+d\Phi_{3},\\
&\textrm{ln}\tau_{2}=b\Phi_{1}+e\Phi_{2}+f\Phi_{3}.\\
\end{aligned}
\end{equation}

Similarly to the previous subsection, canonical normalisation leads to the following conditions:

\begin{equation}
\begin{aligned}
&a^{2}+\frac{3}{2}b^{2}-2ab=\frac{1}{2},\\
&c^{2}+e^{2}+ce=1,\\
&d^{2}+f^{2}+df=1,\\
&a(c+e)=\frac{3}{2}b(c+e),\\
&a(d+f)=\frac{3}{2}b(d+f),\\
&c(d+\frac{1}{2}f)+e(\frac{1}{2}d+f)=0.\\
\end{aligned}
\end{equation}

A solution is:
\begin{equation}
\begin{aligned}
&a=\sqrt{\frac{3}{2}},\\
&b=\sqrt{\frac{2}{3}},\\
&c=e=\sqrt{\frac{1}{3}},\\
&d=-f=1.\\
\end{aligned}
\end{equation}
The Lagrangian then reads:
\begin{equation}
\begin{aligned}
L_{eff}=&\frac{1}{2}(\partial_{u}\Phi_{1})^{2}+\frac{1}{2}(\partial_{u}\Phi_{2})^{2}+\frac{1}{2}(\partial_{u}\Phi_{3})^{2}
+\frac{1}{4}e^{-2\sqrt{\frac{2}{3}}\Phi_{1}-2\Phi_{3}-2\sqrt{\frac{1}{3}}\Phi_{2}}(\partial_{u}a_{1})^{2}\\
&+\frac{1}{4}e^{-2\sqrt{\frac{2}{3}}\Phi_{1}+2\Phi_{3}-2\sqrt{\frac{1}{3}}\Phi_{2}}(\partial_{u}a_{2})^{2}
+\frac{1}{4}e^{\left(2\sqrt{\frac{2}{3}}-2\sqrt{\frac{3}{2}}\right)\Phi_{1}+4\sqrt{\frac{1}{3}}\Phi_{2}}(\partial_{u}a_{3})^{2}-V_{lvs}.\\
\end{aligned}
\end{equation}

Other solutions are also possible but are physically equivalent under field redefinitions.
\section{Type IIA Effective Lagrangian}\label{sec:A}
\subsection{Scalar Potential}
At the minimum, the potential takes the value in (\ref{eq:vmin})
and through the relation  $V_0 = -\frac{3 M_P^2}{R_{AdS}}$ one can express the various couplings as a function of the AdS radius. Surprisingly, all of the dependence on the values of the fluxes drops out, and the the mass matrix is simply given by\footnote{Since the kinetic term (\ref{eq:kt}) is not canonically normalized, we have been careful to use $\tilde{D}= \frac{D}{\sqrt{2}}$.}
\begin{equation}M^2_{ab} = \frac{1}{R^2_{AdS}}
\left(
\begin{array}{cccc}
 34 & 16 & 16 & \,8 \\
 16 & 34 & 16 & \,8 \\
 16 & 16 & 34 & \,8 \\
 \,8 & \,8 & \,8 & 22 \\
\end{array}
\right).
\end{equation}
It can be diagonalised in terms of new fields $ \varphi = U \phi $, where
\begin{equation}
U=\left(
\begin{array}{cccc}
 \,\,2 & \,\,2 & \,\,2 & \,\,1 \\
 -1 & \,\,0 & \,\,0 & \,\,2 \\
 -1 & \,\,0 & \,\,1 & \,\,0 \\
 -1 & \,\,1 & \,\,0 & \,\,0 \\
\end{array}
\right),
\end{equation}
yielding the mass matrix
\begin{equation}
M^2_{D} = \frac{1}{R^2_{AdS}} \text{Diag}(70,18,18,18).
\end{equation}
Using the standard relation $\Delta (\Delta -d) = m^2 R^2_{AdS}$ one obtains\footnote{This equation can have two real solutions, but we only consider the one satisfying the unitarity bound.} \begin{equation}
\Delta_1 = 10 \quad \quad \Delta_2 = \Delta_3 =\Delta_4 =6. 
\end{equation}
It is interesting to notice how the exchange symmetry $\phi_i \leftrightarrow \phi_j$ for the first three fields is only responsible for part of the degeneracy between the conformal dimensions. Indeed, for a generic matrix
\begin{equation}
\renewcommand\arraystretch{0.5}
M = 
\begin{pmatrix}
 a & b & b & c \\
 b & a & b & c \\
 b & b & a & c \\
 c & c & c & d \\
\end{pmatrix}
\end{equation}
with the same symmetry properties only two of the eigenvalues are identical: 
\begin{equation}
\lambda_1 =\lambda_2 =a-b \quad \lambda_{3,4} =\frac{1}{2} \left(a+2 b+d\pm \sqrt{(a+2 b-d)^2+12 c^2}\, \right).
\end{equation}
In order for three (or more) to be equal, the matrix elements must satisfy the additional constraint
\begin{equation}
(a-b-d)b+c^2=0,
\end{equation}
which in the case at hand appears to be coincidental.

Expanding at higher orders, one can obtain the n-point self interactions for the moduli. Given the exponential structure of the potential, derivatives with respect to any of the moduli evaluated at the minimum will be proportional to $V|_{\text{min}}$; hence all vertices will all come with the same factor of $R_{AdS}^{-2}$ in front. This means that also the couplings are independent of the flux choice, and universally fixed apart from an overall constant (corresponding to the $1/N$ expansion parameter) which vanishes in the infinite volume limit.  We report here the cubic terms in the potential, the only ones necessary to compute large-spin anomalous dimensions. 
\begin{equation}
\begin{split}
 &V^{(3)} R^2_{AdS} = 12 \sqrt{2} \tilde{D}^2 \phi _1+12 \sqrt{2} \tilde{D}^2 \phi _2+12 \sqrt{2} \tilde{D}^2 \phi _3+\frac{27 \tilde{D}^3}{\sqrt{2}}+30 \sqrt{2} \tilde{D} \phi _1^2+30 \sqrt{2} \tilde{D} \phi _2^2 \\&+30 \sqrt{2} \tilde{D} \phi _3^2+24 \sqrt{2} \tilde{D} \phi _1 \phi _2+24 \sqrt{2} \tilde{D} \phi _1 \phi _3+24 \sqrt{2} \tilde{D} \phi _2 \phi _3+18 \sqrt{2} \phi _1 \phi _2 \phi _3
\end{split}.
\end{equation} 
In terms of the diagonal fields $\varphi_i$, it becomes
\begin{equation}
\begin{split}
 &V^{(3)} R^2_{AdS} = \frac{\sqrt{2}}{169} \Big(1080\, \varphi _2\varphi_4^2 -360\, \varphi _3 \varphi_4^2 -120\,\varphi_4^3-105\, \varphi _1^2 \varphi_4 -198\, \varphi _2^2 \varphi_4 -360\,\varphi _3^2 \varphi_4 \\ &- 882 \, \varphi _2 \varphi _3\varphi_4+135\, \varphi_1^3+198\, \varphi _2^3-120\, \varphi _3^3+1080\, \varphi _2 \varphi _3^2+315\, \varphi _1^2 \varphi _2-105\, \varphi _1^2 \varphi _3-198\, \varphi _2^2 \varphi _3 \Big)
\end{split}.
\end{equation} 
We notice that, at least up to cubic order, the potential is invariant under the exchange of $\varphi_3$ and $\varphi_4$.
\subsection{Axion Kinetic Term}
The same procedure can be carried out for axions, expanding the lagrangian (\ref{eq:axlag}) about the usual vacuum solution in terms of the canonically normalized fields
\begin{equation}
b'_i = \frac{b_i}{\sqrt{2} \bar{v}_i} \quad \quad \quad  \xi' = \xi e^{\bar{D}}
\end{equation}
and the moduli fluctuations
\begin{equation}
\phi_i' = \phi_i -\bar{\phi}_i \quad \quad D' = D-\bar{D},
\end{equation}
where it is understood that all primes will be omitted in the following.
The resulting expression can be divided into a kinetic term
\begin{equation}\label{eq:kax}
\mathcal{L}_{kin}= \frac{1}{2} e^{-2\sqrt{2} \phi_i} \partial^{\mu} b_i \partial_{\mu} b^i  + \frac{1}{2} e^{\sqrt{2} \tilde{D}} \partial_{\mu} \xi \partial^{\mu} \xi 
\end{equation}
and a potential 
\begin{equation}
\begin{split}
 \mathcal{L}_{\phi b}&= \frac{15}{R^2_{AdS}} e^{2 \sqrt{2} \tilde{D} } \Big[ s_1 b_2 b_3 e^{\sqrt{2}(\phi_1-\phi_2-\phi_3)} + s_2 b_3 b_1 e^{\sqrt{2}(\phi_2-\phi_3-\phi_1)}+ s_3 b_1 b_2 e^{\sqrt{2}(\phi_3-\phi_1-\phi_2)}\Big] \\
& -\frac{9}{2 R^2_{AdS}} e^{ \sqrt{2} (2\tilde{D} -\phi_1-\phi_2-\phi_3)} \Big[ b_1^2+b_2^2+b_3^2+2 s_1s_2 b_1 b_2 +2 s_1s_3 b_1 b_3+2 s_2s_3 b_2 b_3 \Big]\\
& -\frac{12}{ R^2_{AdS}} e^{ \sqrt{2} (2\tilde{D} -\phi_1-\phi_2-\phi_3)} \Big[s_1b_1+s_2b_2+s_3b_3 \Big] \xi -\frac{8}{ R^2_{AdS}} e^{ \sqrt{2} (2\tilde{D} -\phi_1-\phi_2-\phi_3)} \xi^2\\
& -\frac{25}{2 R^2_{AdS}} e^{2 \sqrt{2} \tilde{D} } \Big[ b_1^2 e^{\sqrt{2}(\phi_2+\phi_3-\phi_1)} +b_2^2 e^{\sqrt{2}(\phi_3+\phi_1-\phi_2)} +b_3^2 e^{\sqrt{2}(\phi_1+\phi_2-\phi_3)}  \Big].
\end{split}
\end{equation}
\subsubsection{Spectrum}
Setting all of the moduli fluctuations to zero, this gives the mass matrix
\begin{equation}
M^2_{ab} = \frac{1}{R^2_{AdS}} 
\begin{pmatrix}
34 & 9\, s_{1} s_{2}- 15\,s_{3} & 9\, s_{1} s_{3}- 15\,s_{2} & 12\, s_{1} \\
9\, s_{1} s_{2}- 15\, s_{3} & 34 & 9 \,s_{2} s_{3}- 15 \,s_{1} & 12 s_{2} \\
9 \,s_{1} s_{3}-15\, s_{2} & 9\, s_{2} s_{3}- 15\,s_{1} & 34 & 12\, s_{3} \\
12\, s_{1} & 12 \,s_{2} & 12\, s_{3} & 16
\end{pmatrix},
\end{equation}
as reported in \citep{DeWolfe:2005uu}. Since the matrix depends on the signs $s_i \equiv \text{sgn}(m_0 e_i)$, we can distinguish four different combinations (permutations are indistinguishable):
\paragraph*{Case 1:} s = (1,1,1) 

\noindent  The matrix is diagonalised in term of the new fields $a = U_1 b $, where 
\begin{equation}
U_1 = 
\begin{pmatrix}
 \,\,2 & \,\,0 & \,\,0 & \,\,1 \\
 -1 & \,\,0 & \,\,1 & \,\,0 \\
 -1 & \,\,1 & \,\,0 & \,\,0 \\
 -1 & -1 & -1 & \,\,2 \\
\end{pmatrix}.
\end{equation}
The mass matrix is given by
\begin{equation}
M^2_{D} = \frac{1}{R^2_{AdS}} \text{Diag}(40,40,40,-2),
\end{equation}
resulting in 
\begin{equation}
\Delta_a = (8,8,8,2)  \quad\quad \text{or} \quad\quad \Delta_a = (8,8,8,1).
\end{equation} 

\paragraph*{Case 2:} s = (1,1,-1) 

\noindent  The matrix is diagonalised in term of the new fields $a = U_2 b $, where 
\begin{equation}
U_2 = 
\begin{pmatrix}
 \,\,2 & \,\,2 & -2 & \,\,1 \\
 -1 & \,\,0 & \,\,0 & \,\,2 \\
 \,\,1 & \,\,0 & \,\,1 & \,\,0 \\
 -1 & \,\,1 & \,\,0 & \,\,0 \\
\end{pmatrix}.
\end{equation}
The mass matrix is given by
\begin{equation}
M^2_{D} = \frac{1}{R^2_{AdS}} \text{Diag}(88,10,10,10),
\end{equation}
resulting in 
\begin{equation}
\Delta_a = (11,5,5,5).
\end{equation} 

\paragraph*{Case 3:} s = (1,-1,-1) 

\noindent  The matrix is diagonalised in term of the new fields $a = U_3 b $, where 
\begin{equation}
U_3 = 
\begin{pmatrix}
 \,\,2 & \,\,0 & \,\,0 & \,\,1 \\
 \,\,1 & \,\,0 & 1\,\, & \,\,0 \\
 \,\,1 & \,\,1 & \,\,0 & \,\,0 \\
 -1 & \,\,1 & \,\,1 & \,\,2 \\
\end{pmatrix}.
\end{equation}
The mass matrix is given by
\begin{equation}
M^2_{D} = \frac{1}{R^2_{AdS}} \text{Diag}(40,40,40,-2),
\end{equation}
resulting in 
\begin{equation}
\Delta_a = (8,8,8,2)  \quad\quad \text{or} \quad\quad \Delta_a = (8,8,8,1).
\end{equation} 

\paragraph*{Case 4:} s = (-1,-1,-1) 

\noindent  The matrix is diagonalised in term of the new fields $a = U_4 b $, where 
\begin{equation}
U_4 = 
\begin{pmatrix}
 -2 & -2 & -2 & \,\,1 \\
 \,\,1 & \,\,0 & \,\,0 & \,\,2 \\
 -1 & \,\,0 & \,\,1 & \,\,0 \\
 -1 & \,\,1 & \,\,0 & \,\,0 \\
\end{pmatrix}.
\end{equation}
The mass matrix is given by
\begin{equation}
M^2_{D} = \frac{1}{R^2_{AdS}} \text{Diag}(88,10,10,10),
\end{equation}
resulting in 
\begin{equation}
\Delta_a = (11,5,5,5).
\end{equation} 
\subsubsection{Interactions}
For the interactions, let's start from the ones originating from the kinetic term (\ref{eq:kax}). Again, we can distinguish different depending on the flux signs:
\paragraph*{Case 1:} s = (1,1,1) 
\begin{equation}
\begin{split}
&\mathcal{L}^1_{kin} =
 -\frac{1}{637 \sqrt{2}} \Big[ \\  &(\partial b_1)^2 \big(39 \varphi _1-78  \varphi _2 +26 \varphi _3+26 \varphi _4 \big) + \partial b_1 \partial b_2 \big( 52 \varphi _1 -104  \varphi_2 
+520 \varphi _3-208 \varphi _4 \big)\\
&+ \partial b_1 \partial b_3 \big( 52\varphi _1  -104 \varphi _2 -208\varphi _3  +520  \varphi _4\big)  + \partial b_1 \partial b_4 \big(-60 \varphi_1  - 48  \varphi _2 +16  \varphi _3 +16  \varphi _4 \big)\\
& +(\partial b_2)^2 \big( 148 \varphi _1 -100  \varphi _2 +640  \varphi _3-270  \varphi _4 \big)  +\partial b_2 \partial b_3 \big( -96  \varphi _1 -4 \varphi _2 -120 \varphi _3 -120  \varphi _4\big) \\
&  +\partial b_2 \partial b_4  \big( -40 \varphi _1 -32  \varphi _2  -232  \varphi _3 +132 \varphi _4\big)
+(\partial b_3)^2 \big( 148 \varphi _1 -100  \varphi _2 -270 \varphi _3+640  \varphi _4\big) \\ & +\partial b_3 \partial b_4  \big(-40 \varphi _1-32  \varphi _2 +132 \varphi _3 -232  \varphi _4 \big) +(\partial b_4)^2 \big( 8 \varphi _1-30 \varphi _2+10  \varphi _3+10  \varphi _4 \big) \Big]
\end{split}
\end{equation}
\begin{equation}
\begin{split}
& \mathcal{L}^1_{\varphi b} =
\frac{1}{637} \Big[\\ 
& b_1^2 \Big( (1560 \sqrt{2} -360)\varphi_1+(180-6240 \sqrt{2})\varphi_2+(2080 \sqrt{2} -60)\varphi_3+(2080 \sqrt{2} -60)\varphi_4 \Big) 
\\ & + b_1 b_2 \Big( (2080 \sqrt{2} -480)\varphi_1+(240-8320 \sqrt{2})\varphi_2+(8840 \sqrt{2} -3720)\varphi_3+(1740 - 260 \sqrt{2})\varphi_4 \Big) \\ 
& +b_1 b_3 \Big( (2080 \sqrt{2} -480)\varphi_1+(240-8320 \sqrt{2})\varphi_2+(1740 - 260 \sqrt{2})\varphi_3+(8840 \sqrt{2} -3720)\varphi_4 \Big)
\\&
+b_1 b_4 \Big( (360+120\sqrt{2} )\varphi_1-(180+60 \sqrt{2})\varphi_2+(60+20\sqrt{2})\varphi_3+(60+20\sqrt{2})\varphi_4 \Big) 
\\&
+ b_2^2 \Big( (330-1430 \sqrt{2} )\varphi_1-(165+10205 \sqrt{2})\varphi_2+(10985 \sqrt{2} +1875)\varphi_3-(855 +390\sqrt{2})\varphi_4 \Big) 
 \\&
+b_2 b_3 \Big( (3510\sqrt{2}-810 )\varphi_1+(4050+1885 \sqrt{2})\varphi_2-2145\sqrt{2}\varphi_3-2145\sqrt{2}\varphi_4 \Big) 
\\&
+b_2 b_4 \Big( (240+80\sqrt{2})\varphi_1-(120+40 \sqrt{2})\varphi_2 \Big) +(1860-3020\sqrt{2})\varphi_3+(1530\sqrt{2}-870)\varphi_4\Big) 
\\&
 +b_3^2 \Big( (330-1430 \sqrt{2} )\varphi_1-(165+10205 \sqrt{2})\varphi_2-(855 +390\sqrt{2})\varphi_3+(10985 \sqrt{2} +1875)\varphi_4 \Big) 
 \\&
 +b_3 b_4 \Big( (240+80\sqrt{2})\varphi_1-(120+40 \sqrt{2})\varphi_2 \Big) +(1530\sqrt{2}-870)\varphi_3+(1860-3020\sqrt{2})\varphi_4\Big) 
 \\&
 +b_4^2 \Big( -(90+58 \sqrt{2} )\varphi_1+(45+120 \sqrt{2})\varphi_2-(15 +40\sqrt{2})\varphi_3- (15 +40\sqrt{2})\varphi_4 \Big) 
\Big].
\end{split}
\end{equation}
Both are symmetric under the simultaneous exchanges
\begin{equation}
\varphi_3 \longleftrightarrow \varphi_4 \quad \quad \quad  b_2 \longleftrightarrow b_3.
\end{equation}
\paragraph*{Case 2:} s = (1,-1,1) 
\begin{equation}
\begin{split}
&\mathcal{L}^2_{kin} =
 -\frac{1}{2197 \sqrt{2}} \Big[ \\  &(\partial b_1)^2 \big(47 \varphi _1-30  \varphi _2 +10 \varphi _3+10 \varphi _4 \big) + \partial b_1 \partial b_2 \big( -60 \varphi _1 -48  \varphi_2 
+16 \varphi _3+16 \varphi _4 \big)\\
&+ \partial b_1 \partial b_3 \big( -20\varphi _1  -16 \varphi _2 -896\varphi _3  +456  \varphi _4\big)  + \partial b_1 \partial b_4 \big(20 \varphi_1  +16  \varphi _2 +456 \varphi _3 -896  \varphi _4 \big)\\
& +(\partial b_2)^2 \big( -24 \varphi _1 -222  \varphi _2 +74  \varphi _3+74 \varphi _4 \big)  +\partial b_2 \partial b_3 \big( -16  \varphi _1 -148 \varphi _2 +500 \varphi _3 -176  \varphi _4\big) \\
&  +\partial b_2 \partial b_4  \big( 16 \varphi _1 +148 \varphi _2  +176  \varphi _3 -500 \varphi _4\big)
+(\partial b_3)^2 \big( 448 \varphi _1 -250  \varphi _2 +1210 \varphi _3-480 \varphi _4\big) \\ & +\partial b_3 \partial b_4  \big(456 \varphi _1-176  \varphi _2 +960 \varphi _3 +960  \varphi _4 \big) +(\partial b_4)^2 \big( 448 \varphi _1-250 \varphi _2-480  \varphi _3+1210  \varphi _4 \big) \Big]
\end{split}
\end{equation}
\begin{equation}
\begin{split}
& \mathcal{L}^2_{\varphi b} =
\frac{1}{2197} \Big[\\ 
& b_1^2 \Big( (8 \sqrt{2} +360)\varphi_1-(180+7440 \sqrt{2})\varphi_2+(60 +2480 \sqrt{2})\varphi_3+(60 +2480 \sqrt{2})\varphi_4 \Big) 
\\ & + b_1 b_2 \Big( (2080 \sqrt{2} -480)\varphi_1+(240-8320 \sqrt{2})\varphi_2+(380 \sqrt{2}-60)\varphi_3+(380 \sqrt{2}-60)\varphi_4 \Big) \\ 
& +b_1 b_3 \Big( (760\sqrt{2} -120 )\varphi_1+(60-380 \sqrt{2})\varphi_2-(6780 +11140\sqrt{2})\varphi_3+(3360 +5760 \sqrt{2} )\varphi_4 \Big)
\\&
+b_1 b_4 \Big( (-760\sqrt{2} +120)\varphi_1+(-60+380 \sqrt{2})\varphi_2-(3360 +5760 \sqrt{2} )\varphi_3+(6780 +11140\sqrt{2})\varphi_4 \Big)
\\&
+ b_2^2 \Big( (90+210 \sqrt{2} )\varphi_1-(45+2640 \sqrt{2})\varphi_2+(15 +880\sqrt{2} )\varphi_3+(15+880\sqrt{2})\varphi_4 \Big) 
 \\&
+b_2 b_3 \Big( (60+140\sqrt{2} )\varphi_1-(30+1760 \sqrt{2})\varphi_2+(3390 +6220\sqrt{2})\varphi_3-(1680+2230\sqrt{2})\varphi_4 \Big) 
\\&
+b_2 b_4 \Big( -(60+140\sqrt{2} )\varphi_1+(30+1760 \sqrt{2})\varphi_2+(1680+2230\sqrt{2})\varphi_3-(3390 +6220\sqrt{2})\varphi_4 \Big) 
\\&
 +b_3^2 \Big( -(1680+3920 \sqrt{2} )\varphi_1+(840-5645 \sqrt{2})\varphi_2+(15965\sqrt{2}-13800)\varphi_3+(6480-5160 \sqrt{2})\varphi_4 \Big) 
 \\&
 +b_3 b_4 \Big( -(1710+3990\sqrt{2})\varphi_1+(855-4765 \sqrt{2})\varphi_2 \Big) +(12855\sqrt{2}-15945)\varphi_3+(12855\sqrt{2}-15945)\varphi_4\Big) 
 \\&
  +b_4^2 \Big( -(1680+3920 \sqrt{2} )\varphi_1+(840-5645 \sqrt{2})\varphi_2+(15965\sqrt{2}-13800)\varphi_3+(6480-5160 \sqrt{2})\varphi_4 \Big) 
\Big].
\end{split}
\end{equation}
Both are symmetric under the simultaneous exchanges
\begin{equation}
\varphi_3 \longleftrightarrow \varphi_4 \quad \quad \quad  b_3 \longleftrightarrow -b_4.
\end{equation}
\paragraph*{Case 3:} s = (1,-1,-1)  
\begin{equation}
\mathcal{L}_{kin}^3(b_1,b_2,b_3,b_4) = \mathcal{L}_{kin}^1(b_1,-b_2,-b_3,b_4) 
\end{equation}
\begin{equation}
\mathcal{L}_{\varphi_b}^3(b_1,b_2,b_3,b_4) = \mathcal{L}_{\varphi b}^1(b_1,-b_2,-b_3,b_4) 
\end{equation}
\paragraph*{Case 4:} s = (-1,-1,-1)  
\begin{equation}
\mathcal{L}_{kin}^4(b_1,b_2,b_3,b_4) = \mathcal{L}_{kin}^2(b_1,b_2,b_3,-b_4) 
\end{equation}
\begin{equation}
\mathcal{L}_{\varphi b}^4(b_1,b_2,b_3,b_4) = \mathcal{L}_{\varphi b}^2(b_1,b_2,b_3,-b_4) 
\end{equation}
The relations between cases 1 \& 3 and 2 \& 4 descend from the relations between the matrices $U_1$ \& $U_2$ and $U_3$ \& $U_4$ introduced above.

\input{ArXiv1.bbl}

\end{document}

%% file: ArXiv1.bbl
\providecommand{\href}[2]{#2}\begingroup\raggedright\endgroup